\documentclass[11pt,a4paper]{article}

\usepackage{amsmath,amsfonts,graphicx,natbib,psfrag,color}
\usepackage{algorithm, algorithmic}
\usepackage[textsize=small]{todonotes}
\usepackage{booktabs}

\newcommand{\indep}{\overset{indep}{\sim }}

\title{Augmented pseudo-marginal Metropolis-Hastings for partially observed diffusion processes}

\author{Andrew Golightly$^1$\footnote{andrew.golightly@ncl.ac.uk}, Chris Sherlock$^2$}
\date{\small $^1$ School of Mathematics, Statistics and Physics, Newcastle University, UK\\$^{2}$Department of Mathematics and Statistics, Lancaster University, UK\\}

\begin{document}
\maketitle
\begin{abstract}
We consider the problem of inference for nonlinear, multivariate diffusion processes, 
satisfying It\^o stochastic differential equations (SDEs), using data at discrete times 
that may be incomplete and subject to measurement error. Our starting point is a 
state-of-the-art correlated pseudo-marginal Metropolis-Hastings algorithm, that uses 
correlated particle filters to induce strong and positive correlation between 
successive likelihood estimates. However, unless the measurement error 
or the dimension of the SDE is small, correlation can be eroded by the resampling steps 
in the particle filter. We therefore propose a novel augmentation scheme, that allows 
for conditioning on values of the latent process at the observation times, completely 
avoiding the need for resampling steps. We integrate over the uncertainty at the observation 
times with an additional Gibbs step. Connections between the resulting pseudo-marginal scheme and existing 
inference schemes for diffusion processes are made, {giving a unified inference framework 
that encompasses Gibbs sampling and pseudo marginal schemes.} The methodology is applied in three examples of increasing 
complexity. We find that our approach offers substantial increases in overall efficiency, 
compared to competing methods. 
\end{abstract}


\section{Introduction}
\label{sec:intro}

Although stochastic differential equations (SDEs) have been ubiquitously applied in areas such as finance 
\citep[see e.g.][]{kalogeropoulos10, stramer17}, climate modelling \citep[see e.g.][]{Majda09, Chen14} and 
life sciences \citep[see e.g.][]{GoliWilk11,Fuchs_2013,Wilkinson06,picchini19}, their widespread uptake is 
hindered by the significant challenge of fitting such models to partial data at discrete times. 
General nonlinear, multivariate SDEs rarely admit analytic solutions, necessitating the use of a numerical solution 
(such as that obtained from the Euler-Maruyama scheme). The resulting discretisation error can be controlled 
through the use of intermediate time steps between observation instants. However, integrating over the uncertainty 
at these intermediate times can be computationally expensive.

Upon resorting to discretisation, two approaches to Bayesian inference are apparent. If learning of both the 
static parameters and latent process is required, a Gibbs sampler provides a natural way of exploring the 
joint posterior. The well-studied dependence between the parameters and latent process can be problematic; 
see \cite{StramRobe01} for a discussion of the problem. Gibbs strategies that overcome this dependence have been 
proposed by \cite{StramRobe01} for reducible SDEs and by \cite{GoliWilk08}, \cite{Fuchs_2013}, \cite{Papaspiliopoulos13} and \cite{Meulen17} (among others) for irreducible SDEs. If primary interest lies in learning the parameters, pseudo-marginal Metropolis-Hastings (PMMH) 
schemes \citep{andrieu10,andrieu09b,stramer11,GoliWilk11} can be constructed to directly target the marginal parameter posterior, or, 
with simple modification, the joint posterior over both parameters and the latent process. 
PMMH requires running a particle filter (conditional on a proposed parameter value) at each iteration of the sampler 
to obtain an estimate of the marginal likelihood (henceforth referred to simply as the likelihood). {This can be 
computationally costly, especially in applications involving long time series, since the number of particles 
should be scaled in proportion to the number of data points $n$ to maintain a desired likelihood estimator variance \citep{berard14}}. 
To reduce the variance of the acceptance ratio (for a given number of particles), successive likelihood estimates can be positively correlated 
\citep{dahlin2015,deligiannidis2018}. The resulting correlated PMMH (CPMMH) scheme has been applied in the discretised SDE setting 
by \cite{GoliBrad19}; see also \cite{choppala2016} and \cite{tran2016}. 

As a starting point for this work, we consider a CPMMH scheme. A well known problem with this approach is the use of 
resampling steps in the particle filter, which can destroy correlation between successive likelihood estimates. This problem can 
be alleviated by sorting each particle before propagation using e.g. a Hilbert sorting procedure \citep{deligiannidis2018} 
or simple Euclidean sorting \citep{choppala2016}. However, \cite{GoliBrad19} find that as the level of the noise in the observation process increases, correlation deteriorates. {Moreover, the empirical findings of \cite{deligiannidis2018} suggest that the number of particles should scale as $n^{d/(d+1)}$ (where $d$ is the dimension of the state space of the SDE), so that the advantage of CPMMH over PMMH degrades as the state dimension increases.} Our aim is to avoid resampling altogether. A joint update of the entire latent process would avoid the need for resampling (as implemented by \cite{stramer11} in the PMMH setting) but to be computationally efficient this usually requires an extremely accurate method for sampling the latent process.

Our novel approach is to augment the parameter vector to include the latent process at the observation times, but \emph{not} at the 
intermediate times between observation time instants. Given observations $y$, latent values $x^o$ at observation instants and parameters 
$\theta$, a Gibbs sampler is used to update $\theta$ conditional on $(x^o,y)$, and $x^o$ conditional on $(\theta,y)$. Both steps require 
estimating likelihoods of the form $p(x^o_{t+1}|x^o_t,\theta)$ for which we obtain unbiased estimators via importance sampling. 
Consequently, our approach can be cast within the pseudo-marginal framework and we further use correlation to improve computational efficiency. 
Crucially, we avoid the need for resampling, thus preserving induced positive correlation between likelihood estimates. Furthermore, 
each iteration of our proposed scheme admits steps that are embarrassingly parallel. We refer to the resulting sampler as \emph{augmented 
CPMMH (aCPMMH)}. It should be noted that aCPMMH requires careful initialisation of $x^o$ and subsequently, a suitable proposal mechanism. 
We provide practical advice for initialisation and tuning of proposals for a wide class of SDEs. Special 
cases of aCPMMH are also considered, {and a byproduct of our approach is an inferential framework that encompasses both the pseudo-marginal strategy and Gibbs samplers described in the second paragraph of this section.} In particular, we make clear the connection between aCPMMH and the Gibbs sampler (with reparameterisation) described in \cite{GoliWilk08}. We apply aCPMMH in three examples of increasing complexity and compare against state-of-the-art CPMMH and PMMH schemes. We find that the proposed approach offers an increase in overall efficiency of over an order of magnitude in several settings.     

The remainder of this paper is organised as follows. Background information on the inference problem, PMMH and CPMMH approaches 
is provided in Section~\ref{sec:inf}. Our novel contribution is described in Section~\ref{sec:apmmh} and we explore 
connections between our proposed approach and existing samplers that use data augmentation in Section~\ref{sec:existing}. Applications are 
given in Section~\ref{sec:app}, and conclusions are provided in Section~\ref{sec:disc}, alongside directions for future research. 
 
\section{Bayesian inference via time discretisation}
\label{sec:inf}
Consider a continuous-time $d$-dimensional It\^o process $\{X_t, t\geq t_0\}$
satisfying an SDE of the form
\begin{equation}
dX_t=\alpha(X_t,\theta)\,dt+\sqrt{\beta(X_t,\theta)}\,dW_t,
\quad X_{0}\sim p(x_{0}). \label{eqn:sdmem}
\end{equation}
Here, $\alpha$ is a $d$-vector of drift functions, the 
diffusion coefficient $\beta$ is a $d \times d$ positive 
definite matrix with a square-root representation 
$\sqrt{\beta}$ such that $\sqrt{\beta}\sqrt{\beta}^T=\beta$ 
and $W_t$ is a $d$-vector of (uncorrelated) standard 
Brownian motion processes. We assume that both $\alpha$ 
and $\beta$ depend on $X_t=(X_{1,t},\ldots,X_{d,t})^T$ 
and denote the parameter vector with $\theta=(\theta_{1},\ldots,\theta_{p})^T$ 

Suppose that $\{X_t,t\geq 0\}$ cannot be
observed exactly, but observations 
$y=(y_{1},\ldots,y_{n})^T$ are available on a regular time grid and these
are conditionally independent (given the latent process). We link the
observations to the latent process via an observation model of the form
\begin{equation}\label{eqn:obs}
Y_t = F^T X_t + \epsilon_t, \qquad \epsilon_t|\Sigma\indep N(0,\Sigma),
\end{equation}
where $Y_t$ is a $d_o$-vector, $F$ is a constant $d\times d_o$
matrix and $\epsilon_t$ is a random $d_o$-vector. Note that this setup
allows for only observing a subset of components ($d_o<d$). In settings 
where learning $\Sigma$ is also of interest, the parameter vector $\theta$ 
can be augmented to include the components of $\Sigma$. 

For most problems of interest the form of the SDE in (\ref{eqn:sdmem}) 
will not permit an analytic solution. We therefore work with the
Euler-Maruyama approximation
\[
\Delta X_t\equiv X_{t+\Delta t}-X_t=\alpha(X_t,\theta)\,\Delta t+\sqrt{\beta(X_t,\theta)}\,\Delta W_t
\]
where $\Delta W_t\sim N(0,I_d\Delta t)$. 
To allow arbitrary accuracy of this approximation, we adopt 
a partition of [$t,t+1$]
as
\[
t=\tau_{t,0}<\tau_{t,1}<\tau_{t,2}<\ldots<\tau_{t,m-1}<\tau_{t,m}=t+1
\]
thus introducing $m-1$ intermediate time points with interval widths of length 
\begin{equation}\label{eqn:part}
\Delta\tau\equiv\tau_{t,k+1}-\tau_{t,k}=\frac{1}{m}.
\end{equation}
The Euler-Maruyama approximation is then applied over each interval 
of width $\Delta\tau$, with $m$ chosen by the practitioner, to balance 
accuracy and computational efficiency. The transition density under 
the Euler-Maruyama approximation of $X_{\tau_{t,k+1}}|X_{\tau_{t,k}}=x_{\tau_{t,k}}$ 
is denoted by
\begin{align*}
& p_{e}(x_{\tau_{t,k+1}}|x_{\tau_{t,k}},\theta) \\
&=N\left(x_{\tau_{t,k+1}};x_{\tau_{t,k}}+\alpha(x_{\tau_{t,k}},\theta)\Delta\tau \,,\,\beta(x_{\tau_{t,k}},\theta)\Delta\tau \right)
\end{align*}
where $N(\cdot ; \mu,V)$ denotes the Gaussian density with mean vector $\mu$ and variance matrix $V$.

In what follows, we adopt the shorthand notation 
\[
x_{[t,t+1]} = (x_{\tau_{t,0}},\ldots,x_{\tau_{t,m}})^T
\]
for the 
latent process over the time interval $[t,t+1]$ with an analogous 
notation for intervals of the form $(t,t+1]$ and $(t,t+1)$ 
which ignore $x_{\tau_{t,0}}$ and $(x_{\tau_{t,0}},x_{\tau_{t,m}})$ 
respectively. Hence, the complete latent trajectory is given by
\[
x=(x_{[0,1]}^T,x_{(1,2]}^T,\ldots,x_{(n-1,n]}^T)^T.
\]
The joint density of the latent process over an interval of interest is then given 
by a product of Gaussian densities; for example 
\begin{equation}\label{jointEul}
p_e^{(m)}(x_{(t,t+1]}|x_t,\theta)=\prod_{k=0}^{m-1}p_e(x_{\tau_{t,k+1}}|x_{\tau_{t,k}},\theta)
\end{equation}
{with explicit dependence on the number of intermediate sub-intervals 
made clear via the superscript $(m)$.}

\subsection{Pseudo-marginal Metropolis-Hastings (PMMH)}
\label{sec:pmmh}
Suppose that interest lies in the marginal parameter posterior
\begin{align}
\pi^{(m)}(\theta|y)&\propto \pi(\theta)p^{(m)}(y|\theta) \label{post}
\end{align}
where $\pi(\theta)$ is the prior density ascribed to $\theta$ and 
$p^{(m)}(y|\theta)$ is the (marginal) likelihood under the augmented Euler-Maruyama 
approach. That is,
\begin{align*}
p^{(m)}(y|\theta)&=\int p^{(m)}(x|\theta)p(y|x,\theta)dx
\end{align*}
where 
\begin{align*}
p^{(m)}(x|\theta)&=p(x_{0})\prod_{t=0}^{n-1}p_{e}^{(m)}(x_{(t,t+1]}|x_t,\theta)
\end{align*}
and
\begin{equation}\label{obsdens}
p(y|x,\theta)=\prod_{t=1}^{n}p(y_{t}|x_{t},\theta).
\end{equation}
Although $\pi^{(m)}(\theta|y)$ is typically complicated by the intractable likelihood term, 
$p^{(m)}(y|\theta)$, the latter can be unbiasedly estimated using a particle filter \citep{delmoral04,pitt12}. 
We write the estimator as $\hat{p}_U^{(m)}(y|\theta)$ with explicit dependence on $U\sim p(u)$, that is, the collection 
of all random variables of which a realisation $u$ gives the estimate $\hat{p}_u^{(m)}(y|\theta)$. Algorithm~\ref{PF} 
gives the necessary steps for the generation of $\hat{p}_u^{(m)}(y|\theta)$, with the explicit role of $u$ suppressed 
for simplicity. A key requirement of the particle filter is the ability to simulate latent trajectories $x_{(t,t+1]}$ 
at each time $t$. To yield a reasonable particle weight, such trajectories must be consistent with $x_t$ and $y_{t+1}$ 
and are typically termed \emph{bridges}. In this article we generate bridges by drawing from a density of the form
\[
g\left(x_{(t,t+1]}|x_{t},y_{t+1},\theta\right)=\prod_{k=0}^{m-1}g(x_{\tau_{t,k+1}}|x_{\tau_{t,k}},y_{t+1},\theta)
\]
where the constituent terms take the form
\begin{align}
 g(x_{\tau_{t,k+1}}|x_{\tau_{t,k}},y_{t+1},\theta) & =N\left(x_{\tau_{t,k+1}};x_{\tau_{t,k}}+\mu(x_{\tau_{t,k}},y_{t+1},\theta)\Delta\tau\,,\, \Psi(x_{\tau_{t,k}},\theta)\Delta\tau\right) \label{bridge}
\end{align}
for suitable choices of $\mu(x_{\tau_{t,k}},y_{t+1},\theta)$ and $\Psi(x_{\tau_{t,k}},\theta)$. The form of (\ref{bridge}) permits a wide choice 
of bridge construct and we refer the reader to \cite{whitaker2017} and \cite{schauer17} for several options. Throughout this paper, we take 
\begin{align}\label{eqn:mdb1}
 \mu(x_{\tau_{t,k}},y_{t+1},\theta) &=\alpha_{k}+\beta_{k} F\left(F^T\beta_{k}F\Delta_k + \Sigma\right)^{-1}\left\{y_{t+1}-F^T(x_{\tau_{t,k}}+\alpha_{k}\Delta_k)\right\}   
\end{align}
and
\begin{equation}\label{eqn:mdb2}
\Psi(x_{\tau_{t,k}},\theta)=\beta_{k}-\beta_{k} F\left(F^T\beta_{k} F\Delta_k + \Sigma\right)^{-1}F^T\beta_{k}\Delta\tau 
\end{equation}
where $\Delta_{k}=t+1-\tau_{t,k}$ and we adopt the shorthand notation that $\alpha_{k}:=\alpha(x_{\tau_{t,k}},\theta)$ 
and $\beta_k:=\beta(x_{\tau_{t,k}},\theta)$. We note that (\ref{eqn:mdb1}) and (\ref{eqn:mdb2}) correspond to the 
(extension to partial and noisy observations of the) modified diffusion bridge construct of \cite{DurhGall02}. 
We may write the construct generatively as 
\begin{equation}\label{bridgeGen}
x_{\tau_{t,k+1}}=x_{\tau_{t,k}}+\mu(x_{\tau_{t,k}},y_{t+1},\theta)\Delta\tau + \sqrt{\Psi(x_{\tau_{t,k}},\theta)\Delta\tau}\,u_{\tau_{t,k}}
\end{equation}
where $u_{\tau_{t,k}}\sim N(0,I_d)$. It should then be clear that an estimate of the likelihood, $\hat{p}_u^{(m)}(y|\theta)$, 
is a deterministic function of the Gaussian innovations driving the bridge construct, and additionally, any random variates required in the 
resampling step of Algorithm~\ref{PF}. We use systematic resampling, which requires a single uniform innovation per resampling step.

\begin{algorithm}[t]
\caption{Particle filter}\label{PF}
\textbf{Input:} observations $y=(y_1,\ldots,y_n)^T$, parameter $\theta$, auxiliary variable $u$ and the number of particles $N$.
\begin{enumerate}
\item \textbf{Initialise}. For $i=1,\ldots,N$ sample particle $x_0^{i}$ from the initial state distribution and assign weight $w_0^{i}=1/N$. 
\item For times $t=0,1,\ldots ,n-1$:
\begin{itemize}
\item[(a)] \textbf{Resample.} For $i=1,\ldots,N$, sample the index $a_{t}^{i}\sim \mathcal{M}\big(w_{t}^{1:N}\big)$ of the ancestor of particle $i$, where $\mathcal{M}(w_t^{1:N})$ denotes a categorical distribution on $\{1,\ldots,N\}$ with probabilities $w_t^{1:N}$.
\item[(b)] \textbf{Propagate.} Draw $x_{(t,t+1]}^{i}\sim g\big(\cdot|x_{t}^{a_{t}^{i}},y_{t+1},\theta\big)$, $i=1,\ldots,N$.
\item[(c)] \textbf{Compute weights.} For $i=1,\ldots,N$ set
\begin{align*}
\tilde{w}_{t+1}^{i}&=\frac{p(y_{t+1}|x_{t+1}^{i},\theta)p_e^{(m)}\big(x_{(t,t+1]}^{i}|x_{t}^{a_{t}^{i}},\theta\big)}
{g\big(x_{(t,t+1]}^{i}|x_{t}^{a_{t}^{i}},y_{t+1},\theta\big)}, \\
w_{t+1}^{i}&=\frac{\tilde{w}_{t+1}^{i}}{\sum_{j=1}^{N}\tilde{w}_{t+1}^{j}}.
\end{align*}
\end{itemize}
\end{enumerate}
\textbf{Output:} estimate $\hat{p}_{u}^{(m)}({y}|\theta)=N^{-n}\prod_{t=0}^{n-1}\sum_{i=1}^{N}\tilde{w}_{t+1}^{i}$ of the observed data likelihood.
\end{algorithm}   

The pseudo-marginal Metropolis-Hastings (PMMH) scheme \citep{andrieu09b,andrieu09} is a Metropolis-Hastings (MH) scheme targeting the joint density
\[
\pi^{(m)}(\theta,u|y)\propto \pi(\theta)\hat{p}_{u}^{(m)}(y|\theta)p(u)
\]
for which it is easily checked (using $\int \hat{p}_{u}^{(m)}(y|\theta)p(u)du=p^{(m)}(y|\theta)$) that $\pi^{(m)}(\theta|y)$ is a marginal density. 
Hence, for a proposal density that factorises as $q(\theta'|\theta)q(u')$, the MH acceptance probability is 
\begin{equation}\label{aprob}
\alpha (\theta',u'|\theta,u)=\min\left\{1\,,\,\frac{\pi(\theta')}{\pi(\theta)}\times\frac{\hat{p}_{u'}^{(m)}(y|\theta')}{\hat{p}_{u}^{(m)}(y|\theta)}\times\frac{q(\theta|\theta')}{q(\theta'|\theta)}\right\}.
\end{equation}
The variance of $\hat{p}_{u}^{(m)}(y|\theta)$ is controlled by the number of particles $N$, which should be chosen to balance both mixing 
and computational efficiency. For example, as the variance of the likelihood estimator increases, the acceptance probability of 
the pseudo-marginal MH scheme decreases to 0 \citep{pitt12}. Increasing $N$ results in more acceptances at increased computational cost. 
Practical advice for choosing $N$ is given by  \cite{Sherlock2015} and \cite{Doucet2015} under two different sets of simplifying assumptions. 
Given a parameter value with good support under the posterior (e.g. the marginal posterior mean, estimated from a pilot run), 
we select $N$ so that the estimated log-likelihood at this parameter value has a standard deviation of roughly 1.5. Unfortunately, 
the value of $N$ required to meet this condition is often found to be impractically large. Therefore, we consider a variance reduction 
technique which is key to our proposed approach.

\subsection{Correlated pseudo-marginal Metropolis-Hastings (CPMMH)}
\label{sec:cpmmh}

The correlated pseudo-marginal scheme \citep{deligiannidis2018,dahlin2015} aims to reduce the variance of the 
acceptance ratio in (\ref{aprob}) by inducing strong and positive correlation between successive estimates 
of the observed data likelihood in the MH scheme. This can be achieved by taking a proposal 
$q(\theta'|\theta)K(u'|u)$ where $K(u'|u)$ satisfies the detailed balance equation
\[
K(u'|u)p(u)=K(u|u')p(u').
\]
Recall that $u$ consists of the collection of Gaussian random variates 
used to propagate the state particles (2(b) in Algorithm~\ref{PF}) 
and any variates required in the resampling step 
(2(a) in Algorithm~\ref{PF}). The Uniform random variate required for 
systematic resampling can be obtained by applying the inverse Gaussian cdf to a Gaussian 
draw. Hence, $u$ consists entirely of standard Gaussian variates and it is then natural 
to set
\begin{equation}\label{cn}
K(u'|u)=\textrm{N}\left(u';\,\rho u\,,\,\left(1-\rho^2\right)I_{d^*}\right)
\end{equation}   
where $d^*$ is the total number of required innovations. 
We note that the density in (\ref{cn}) corresponds to a Crank--Nicolson proposal 
density for which it is easily checked that $p(u)=N(u;0,I_{d^*})$ is invariant. The parameter 
$\rho$ is chosen by the practitioner, with $\rho\approx 1$ inducing strong and positive 
correlation between $\hat{p}_{u'}^{(m)}(y|\theta')$ and $\hat{p}_{u}^{(m)}(y|\theta)$. The 
correlated pseudo-marginal Metropolis-Hastings scheme is given by Algorithm~\ref{algcor}, which 
should be used in conjunction with a modified version of Algorithm~\ref{PF} to induce 
the desired correlation. However, 
the resampling step has the effect of breaking down correlation between successive likelihood estimates. 
To alleviate this problem, the particles can be sorted immediately after propagation 
e.g.\ using a Hilbert sorting procedure \citep{deligiannidis2018} or 
simple Euclidean sorting \citep{choppala2016}. Given a distance metric between particles, 
the particles are sorted as follows: the first particle in the sorted list is the one which 
has the smallest first component; for $j=2,\dots,N$, the $j$th particle in the sorted list 
is chosen to be the one among the unsorted $N-j+1$ particles that is closest to the $j-1$th sorted particle.

Upon choosing a value of $\rho$ (e.g. $\rho=0.99$), the number of particles $N$ can be chosen 
to minimise the distance between successive log estimates of marginal likelihood \citep{deligiannidis2018}. 
In practice, we choose $N$ so that the variance of the logarithm of the ratio 
$\hat{p}_{u'}^{(m)}(y|\theta) / \hat{p}_{u}^{(m)}(y|\theta)$ 
is around 1, for $\theta$ set at some central posterior value.
 
\begin{algorithm}[t]
\caption{Correlated PMMH scheme (CPMMH)}\label{algcor}
\textbf{Input:} parameter value $\theta^{(0)}$, correlation parameter $\rho$ and the number of CPMMH iterations $n_{\textrm{iters}}$.
\begin{enumerate}
\item \textbf{Initialise.} Draw $u^{(0)}\sim p(\cdot)$ and compute $\hat{p}_{u^{(0)}}^{(m)}({y}|\theta^{(0)})$ by running Algorithm~\ref{PF} with $(\theta,u)=(\theta^{(0)},u^{(0)})$. Set the iteration counter $i=1$.
\item \textbf{Update parameters.}
\begin{itemize}
\item[(a)] Draw $\theta'\sim q(\cdot | \theta^{(i-1)})$ and $u'\sim K(\cdot|u^{(i-1)})$.
\item[(b)] Compute $\hat{p}_{u'}^{(m)}({y}|\theta')$ by running Algorithm~\ref{PF} with $(\theta,u)=(\theta',u')$.
\item[(c)] With probability $\alpha (\theta',u'|\theta^{(i-1)},u^{(i-1)})$ given by (\ref{aprob}), 
put $(\theta^{(i)},u^{(i)})=(\theta',u')$ otherwise store the current values $(\theta^{(i)},u^{(i)})=(\theta^{(i-1)},u^{(i-1)})$.
\end{itemize}
\item If $i=n_{\textrm{iters}}$, stop. Otherwise, set $i:=i+1$ and go to step 2.
\end{enumerate}
\textbf{Output:} $\theta^{(1)},\ldots,\theta^{(n_{\textrm{iters}})}$.
\end{algorithm}

{There are a nuber of limitations regarding the implementation of CPMMH as described here, 
which motivate the approach of Section~\ref{sec:apmmh}.} 
Although sorting particle trajectories after propagation can in theory alleviate the effect of resampling 
on maintaining correlation between successive likelihood estimates, the sorting procedure can be 
unsatisfactory in practice. For example, the Euclidean sorting procedure described above (and 
implemented within the SDE context in \cite{GoliBrad19}) {sorts trajectories $x_{(t,t+)]}^{1:N}$ 
between observation times by applying the procedure to the particle states $x_{t+1}^{1:N}$ at the observation 
times. Consequently, trajectories with similar values at time $t+1$ may exhibit qualitatively 
different behaviour between observation times, leading to potentially significant differences 
in likelihood contributions (via the particle weights). In turn, this may break down correlation 
between likelihood values at successive iterations.} The procedure is therefore likely to be 
particularly ineffectual when the measurement error variance is large 
relative to stochasticity inherent in the latent diffusion process, {or, as the dimension of 
the SDE increases.} Resampling may be executed less often, although choosing a resampling schedule \emph{a priori} 
may necessarily be \emph{ad hoc}. {Moreover, reducing the number of resampling steps, 
or indeed omitting the resampling step altogether (so that an importance sampler is obtained) would necessitate 
a bridge construct that samples over the entire inter-observation interval from an approximation 
that is very close to the true (but intractable) conditioned diffusion process,} otherwise the resulting importance sampler weights are likely to have high variance. 
In what follows, we derive a novel approach which avoids resampling altogether, without recourse to 
importance sampling of the entire latent process. 

\section{Augmented CPMMH (aCPMMH)}
\label{sec:apmmh}


It will be helpful throughout this section to denote $x^o$ as the values of $x$ at the observation 
times $1,2,\ldots,n$, and $x^L$ as the values of $x$ at the remaining intermediate times. That is
\[
x^L=(x_{[0,1)}^T,x_{(1,2)}^T,\ldots,x_{(n-1,n)}^T)^T.
\]
It is also possible to treat $x_n$ as a latent variable. In what follows, 
we include $x_n$ in $x^o$ for ease of exposition. 
  
Rather than target the posterior in (\ref{post}), we target the joint posterior
\begin{equation}\label{post2}
\pi^{(m)}(\theta,x^o|y)\propto \pi(\theta)p^{(m)}(x^o|\theta) p(y|x^o,\theta) 
\end{equation}
where 
\begin{equation}\label{like}
p^{(m)}(x^o|\theta)=\int p^{(m)}(x|\theta) dx^L
\end{equation}
and $p(y|x^o,\theta)=p(y|x,\theta)$ as in (\ref{obsdens}). Although the integral in (\ref{like}) 
will be intractable, we may estimate it unbiasedly as follows. 

\subsection{Sequential importance sampling}
\label{sec:apmmhSIS}
We adopt the factorisation
\[
p^{(m)}(x^o|\theta)=p^{(m)}(x_1|\theta)\prod_{t=1}^{n-1}p^{(m)}(x_{t+1}|x_{t},\theta)
\]
and note that the constituent terms can be written as
\begin{align}\label{trans}
p^{(m)}(x_1|\theta)&=\int p(x_0) p_{e}^{(m)}(x_{(0,1]}|x_0,\theta)dx_{[0,1)}, \nonumber \\  
p^{(m)}(x_{t+1}|x_{t},\theta)&=\int p_{e}^{(m)}(x_{(t,t+1]}|x_t,\theta)dx_{(t,t+1)};
\end{align}
recall that $p_{e}^{(m)}(x_{(t,t+1]}|x_t,\theta)$ is given by (\ref{jointEul}). Now, given some suitable importance density\\ 
$g(x_{(t,t+1)}|x_t,x_{t+1},\theta)$, we may write
\begin{align*}
p^{(m)}(x_{t+1}|x_{t},\theta) &=\int \frac{p_{e}^{(m)}(x_{(t,t+1]}|x_t,\theta)}{g(x_{(t,t+1)}|x_t,x_{t+1},\theta)}g(x_{(t,t+1)}|x_t,x_{t+1},\theta)dx_{(t,t+1)}\\
&=E_{x_{(t,t+1)}\sim g}\left\{\frac{p_{e}^{(m)}(x_{(t,t+1]}|x_t,\theta)}{g(x_{(t,t+1)}|x_t,x_{t+1},\theta)}\right\},
\end{align*}
and a similar expression can be obtained for $p^{(m)}(x_1|\theta)$. Hence, given $N$ draws $x_{(t,t+1)}^i$, $i=1,\ldots,N$ 
from the density $g(\cdot|x_t,x_{t+1},\theta)$, 
a realisation of an unbiased estimator of $p^{(m)}(x_{t+1}|x_{t},\theta)$ is
\begin{equation}\label{transEst}
\hat{p}^{(m)}_{u_t}(x_{t+1}|x_{t},\theta)=\frac{1}{N}\sum_{i=1}^{N}\frac{p_{e}^{(m)}(x_{(t,t+1]}^i|x_t,\theta)}{g(x_{(t,t+1)}^i|x_t,x_{t+1},\theta)}
\end{equation}
with the convention that $x_{t+1}^i=x_{t+1}$ for all $i$. We recognise (\ref{transEst}) as an importance sampling estimator of (\ref{trans}). 
An unbiased importance sampling estimator of $p^{(m)}(x_1|\theta)$ can be obtained in a similar manner, by using an importance density $p(x_0)g(x_{(0,1)}|x_0,x_{1},\theta)$.
 
We take $g(x_{(t,t+1)}|x_t,x_{t+1},\theta)$ as a simplification of the bridge construct used in Section~\ref{sec:pmmh} so that
\[
g\left(x_{(t,t+1)}|x_{t},x_{t+1},\theta\right)=\prod_{k=0}^{m-2}g(x_{\tau_{t,k+1}}|x_{\tau_{t,k}},x_{t+1},\theta)
\]
where $g\left(x_{(t,t+1)}|x_{t},x_{t+1},\theta\right)$ has the form \eqref{bridge} but with the exact $x_{t+1}$ taking the place of the noisy $y_{t+1}$. Since $\Sigma=0$, \eqref{eqn:mdb1} and \eqref{eqn:mdb2} simplify to
\begin{align}\label{bridge2}
\mu(x_{\tau_{t,k}},x_{t+1})&=\frac{x_{t+1}-x_{\tau_{t,k}}}{t+1-\tau_{t,k}},\qquad \Psi(x_{\tau_{t,k}},\theta)=\frac{t+1-\tau_{t,k+1}}{t+1-\tau_{t,k}}\beta(x_{\tau_{t,k}},\theta).
\end{align}
We make clear the role of the of the innovation vector $u_t=(u_{t,0},\ldots,u_{t,m-2})^T$ in (\ref{transEst}) by writing the bridge construct generatively as in (\ref{bridgeGen}) but with $\mu$ and $\Psi$ given by (\ref{bridge2}).

Now, since the $x_{(t,t+1)}$, $t=0,\ldots,n-1$ are conditionally independent given $x^o$, we may unbiasedly estimate
$p^{(m)}(x^o|\theta)$ with
\begin{equation}\label{margll}
\hat{p}_U^{(m)}(x^o|\theta)=p^{(m)}_{U_0}(x_1|\theta)\prod_{t=1}^{n-1}\hat{p}_{U_t}^{(m)}(x_{t+1}|x_{t},\theta),
\end{equation}
realisations of which may be computed by running the importance sampler in Algorithm~\ref{imp} for each $t=0,\ldots,n-1$. 
Note that each $t$-iteration of Algorithm~\ref{imp} can be performed in parallel if desired.

\begin{algorithm}[t]
\caption{Importance sampling}\label{imp}
\textbf{Input:} parameter $\theta$, latent values $x_{t},x_{t+1}$, auxiliary variable $u_{t}$ and the number of importance samples $N$.
\begin{itemize}
\item[(a)] \textbf{Sample.} Draw $x_{(t,t+1)}^{i}\sim g\big(\cdot|x_{t},x_{t+1},\theta\big)$, $i=1,\ldots,N$.\\
(If $t=0$, draw $x_0^i \sim p(\cdot)$ and $x_{(0,1)}^{i}\sim g\big(\cdot|x_{0}^i,x_{1},\theta\big)$, $i=1,\ldots,N$.)
\item[(b)] \textbf{Compute weights.} For $i=1,\ldots,N$ set
\[
\tilde{w}_{t+1}^{i}=\frac{p_e^{(m)}\big(x_{(t,t+1]}^{i}|x_{t},\theta\big)}
{g\big(x_{(t,t+1)}^{i}|x_{t},x_{t+1},\theta\big)} 
\]
\end{itemize}
\textbf{Output:} estimate $\hat{p}_{u_{t}}^{(m)}(x_{t+1}|x_{t},\theta)=\frac{1}{N}\sum_{i=1}^{N}\tilde{w}_{t+1}^{i}$ of $p^{(m)}(x_{t+1}|x_t,\theta)$ (or $\hat{p}_{u_{0}}^{(m)}(x_{1}|\theta)=\frac{1}{N}\sum_{i=1}^{N}\tilde{w}_{1}^{i}$ of $p^{(m)}(x_{1}|\theta)$ if $t=0$).
\end{algorithm}   

\subsection{Algorithm}
\label{sec:apmmhAlg}

We adopt a pseudo-marginal approach by targeting the joint density
\begin{equation}\label{pmpost}
\pi^{(m)}(\theta,x^o,u|y)\propto \pi(\theta)\hat{p}_{u}^{(m)}(x^o|\theta)p(y|x^o,\theta)p(u)
\end{equation}
for which it is easily checked that the posterior of interest, $\pi^{(m)}(\theta,x^o|y)$ given by 
(\ref{post2}), is a marginal density. The form of (\ref{pmpost}) immediately suggests a Gibbs sampler 
which alternates between draws from the full conditional densities (FCDs)
\newpage
\begin{enumerate}
\item $\pi^{(m)}(\theta|u,x^o,y)\propto \pi(\theta)\hat{p}_{u}^{(m)}(x^o|\theta)p(y|x^o,\theta)$
\item $\pi^{(m)}(x^o,u|\theta,y)\propto \hat{p}_{u}^{(m)}(x^o|\theta)p(y|x^o,\theta)p(u)$. 
\end{enumerate}
{Hence, unlike the (C)PMMH scheme, the latent process at the observation times is no longer integrated out. 
Nevertheless, the sampler targets a posterior for which the latent process between observation instants 
is marginalised over, and this is crucial for side-stepping the well known dependence problem between the parameters 
and latent process. Note that as the number of importance samples $N\to\infty$, the scheme can be seen as an idealised 
Gibbs sampler that alternates between draws of $\pi^{(m)}(\theta|x^o,y)$ and $\pi^{(m)}(x^o|\theta,y)$. For $N=1$, 
the scheme is an extension of the modified innovation scheme of \cite{GoliWilk08}, as discussed further in 
Section~\ref{sec:existing}.}

Metropolis-within-Gibbs steps are necessary for generating draws from the FCDs above. 
To sample the FCD $\pi^{(m)}(\theta|u,x^o,y)$ we use a proposal density 
$q(\theta'|\theta)$ so that the acceptance probability is given by
\begin{align}\label{aprob2}
\alpha (\theta'|\theta,u,x^o)&=\min\left\{1\,,\,\frac{\pi(\theta')}{\pi(\theta)}\times\frac{\hat{p}_{u}^{(m)}(x^o|\theta')}{\hat{p}_{u}^{(m)}(x^o|\theta)}\times\frac{p(y|x^o,\theta')}{p(y|x^o,\theta)}
\times\frac{q(\theta|\theta')}{q(\theta'|\theta)}\right\}.
\end{align}

Given that $\pi^{(m)}(x^o,u|\theta,y)$ may be high dimensional, we propose to update $(x^o,u)$ in separate blocks 
corresponding to each time component of $x^o$. For $t=1,\ldots,n-1$ we have that
\begin{align*}
\pi^{(m)}(x_t,u_{t-1},u_t|x_{t-1},x_{t+1},y_t,\theta) &\propto \hat{p}_{u_{t-1}}^{(m)}(x_t|x_{t-1},\theta)\hat{p}_{u_t}^{(m)}(x_{t+1}|x_{t},\theta)p(y_t|x_t,\theta)p(u_{t-1})p(u_t)
\end{align*}
where, for example $p(u_t)=N(u_t;0,I_{N(m-1)d})$ {(assuming $N$ importance samples, $(m-1)$ 
intermediate time points between observation instants and a $d$-dimensional latent process)}. 
For $t=1$, $\hat{p}_{u_{t-1}}^{(m)}(x_t|x_{t-1},\theta)$ is replaced by $\hat{p}_{u_{0}}^{(m)}(x_1|\theta)$. The full conditionals for the remaining end-point is given by
\begin{align*}
\pi^{(m)}(x_n,u_{n-1}|x_{n-1},y_n,\theta) &\propto \hat{p}_{u_{n-1}}^{(m)}(x_n|x_{n-1},\theta)p(y_n|x_n,\theta)p(u_{n-1}).
\end{align*}
We sample from each FCD using a Metropolis-within-Gibbs step. For each $t=1,\ldots,n-1$ 
we use a proposal density of the form 
\[
q(x_t',u_{(t-1,t)}'|x_t,u_{(t-1,t)})=q(x_t'|x_t)K_2(u_{(t-1,t)}'|u_{(t-1,t)})
\]
where $K_2(u'_{(t-1,t)}|u_{(t-1,t)})=K(u'_{t-1}|u_{t-1})K(u'_t|u_t)$ {and we use the shorthand notation 
$u_{(t-1,t)}=(u_{t-1},u_t)$}. Hence, the innovations are updated using a Crank-Nicolson kernel. 
The end-point proposal is defined similarly, with 
$q(x_n',u_{n-1}'|x_n,u_{n-1})=q(x_n'|x_n)K(u_{n-1}'|u_{n-1})$. We recall that the Crank-Nicolson kernel 
satisfies detailed balance with respect to the innovation density to arrive at the acceptance probabilities
\begin{align}
&\alpha(x_t',u_{(t-1,t)}'|x_t,u_{(t-1,t)},x_{t-1},x_{t+1},\theta) \nonumber \\
&=\min\left\{ 1\,,\,\frac{\hat{p}_{u_{t-1}'}^{(m)}(x_t'|x_{t-1},\theta)}{\hat{p}_{u_{t-1}}^{(m)}(x_t|x_{t-1},\theta)}\times 
\frac{\hat{p}_{u_t'}^{(m)}(x_{t+1}|x_{t}',\theta)}{\hat{p}_{u_t}^{(m)}(x_{t+1}|x_{t},\theta)} \quad\times \frac{p(y_t|x_t',\theta)}{p(y_t|x_t,\theta)} \times \frac{q(x_t | x_t')}{q(x_t' | x_t)}\right\}\label{aprobx1}
\end{align} 
for $t=1,\ldots,n-1$. For the end-point update, the acceptance probability is
\begin{align}
\alpha(x_n',u_{n-1}'|x_n,u_{n-1}',x_{n-1},\theta)&=\min\left\{1\,,\,\frac{\hat{p}_{u_{n-1}'}^{(m)}(x_n'|x_{n-1},\theta)}{\hat{p}_{u_{n-1}}^{(m)}(x_n|x_{n-1},\theta)}\times 
\frac{p(y_n|x_n',\theta)}{p(y_n|x_n,\theta)} 
\times \frac{q(x_n | x_n')}{q(x_n' | x_n)}\right\}.\label{aprobx3}
\end{align}
It is evident that the innovations $(u_1,\ldots,u_{n-1})$ are updated twice per Gibbs iteration. We note that a scheme that only updates these innovations once per Gibbs iteration is also possible, but eschew this approach in favour of the above, which promotes better exploration of the innovation variable space. Further tuning considerations are discussed in Section~\ref{sec:init}. 

We refer to the resulting inference scheme as augmented CPMMH (aCPMMH). The scheme is summarised by 
Algorithm~\ref{algaug}. Note that the components of the latent process $X^o$ at the observation times 
{and auxiliary variables $u$} are updated in steps 3-5; step 3 updates $(x_t,u_{(t-1,t)})$ for $t=1,3,\ldots,n-1$ and step 4 for $t=2,4,\ldots,n-2$ (assuming, WLOG, that $n$ is even). Step 5 updates the final value $(x_n,u_{n-1})$. Updating in this way allows for embarrisingly parallel operations over $t$ (at steps 2, 3 and 4). {Note that, as presented, uncertainty for the initial value $x_0$ is integrated over as part of the importance sampler (Algorithm~\ref{imp}). If required, aCPMMH can be modified either to treat $x_0$ as part of the parameter vector $\theta$ or with an extra step that updates $x_0$ (and therefore $u_0$) conditional on $x_1$ and $\theta$.} 

\begin{algorithm}[ht!]
\caption{Augmented CPMMH scheme (aCPMMH)}\label{algaug}
\textbf{Input:} parameter and latent values $(\theta^{(0)},x^{o,(0)})$, correlation parameter $\rho$, number of importance samples $N$ and the number of iterations $n_{\textrm{iters}}$.
\begin{enumerate}
\item \textbf{Initialise.} Draw $u^{(0)}\sim p(\cdot)$ and compute $\hat{p}_{u^{(0)}}^{(m)}({x^{o,(0)}}|\theta^{(0)})$ by running Algorithm~\ref{imp} for $t=0,\ldots,n-1$. Set the iteration counter $i=1$.
\item \textbf{Update parameters $\boldsymbol{\theta}$.}
\begin{itemize}
\item[(a)] Draw $\theta'\sim q(\cdot | \theta^{(i-1)})$ and compute 
$\hat{p}_{u^{(i-1)}}^{(m)}(x^{o,(i-1)}|\theta')$ by running Algorithm~\ref{imp} for $t=0,\ldots,n-1$. 
\item[(b)] With probability $\alpha (\theta'|\theta^{(i-1)},u^{(i-1)},x^{o,(i-1)})$ given by (\ref{aprob2}) 
put $\theta^{(i)}=\theta'$ otherwise store the current value $\theta^{(i)}=\theta^{(i-1)}$.
\end{itemize}
\item \textbf{Update $\boldsymbol{(x_t,u_{(t-1,t)})}$, $\boldsymbol{t=1,3,\ldots,n-1}$.}
\begin{itemize}
\item[(a)] Draw $x_t'\sim q(\cdot | x_t^{(i-1)})$ and $u_{(t-1,t)}'\sim K_2(\cdot|u_{(t-1,t)}^{(i-1)})$. 
Compute $\hat{p}_{u_{t-1}'}^{(m)}(x'_t|x^{(i-1)}_{t-1},\theta^{(i)})$ and $\hat{p}_{u_{t}'}^{(m)}(x^{(i-1)}_{t+1}|x'_t,\theta^{(i)})$  
by running iterations $t-1$ and $t$ of Algorithm~\ref{imp}.
\item[(b)] With probability\\ $\alpha(x_t',u_{(t-1,t)}'|x_t^{(i-1)},u_{(t-1,t)}^{(i-1)},x_{t-1}^{(i-1)},x_{t+1}^{(i-1)},\theta^{(i)})$ given by (\ref{aprobx1}) 
put $x^{(i)}_t=x'_t$ and $u_{(t-1,t)}^{(i)}=u_{(t-1,t)}'$ otherwise store the current value $x^{(i)}_t=x^{(i-1)}_t$ and $u_{(t-1,t)}^{(i)}=u_{(t-1,t)}^{(i-1)}$. 
\end{itemize}
\item \textbf{Update $\boldsymbol{(x_t,u_{(t-1,t)})}$, $\boldsymbol{t=2,4,\ldots,n-2}$.}
\begin{itemize}
\item[(a)] Draw $x_t'\sim q(\cdot | x_t^{(i-1)})$ and $u_{(t-1,t)}'\sim K_2(\cdot|u_{(t-1,t)}^{(i)})$. 
Compute $\hat{p}_{u_{t-1}'}^{(m)}(x'_t|x^{(i)}_{t-1},\theta^{(i)})$ and $\hat{p}_{u_{t}'}^{(m)}(x^{(i)}_{t+1}|x'_t,\theta^{(i)})$  
by running iterations $t-1$ and $t$ of Algorithm~\ref{imp}.
\item[(b)] With probability\\ $\alpha(x_t',u_{(t-1,t)}'|x_t^{(i-1)},u_{(t-1,t)}^{(i)},x_{t-1}^{(i)},x_{t+1}^{(i)},\theta^{(i)})$ given by (\ref{aprobx1}) 
put $x^{(i)}_t=x'_t$ and $u_{(t-1,t)}^{(i)}=u_{(t-1,t)}'$ otherwise store the current value $x^{(i)}_t=x^{(i-1)}_t$ (and $u_{(t-1,t)}^{(i)}$ remains unchanged). 
\end{itemize}
\item \textbf{Update $\boldsymbol{(x_n,u_{n-1})}$.}
\begin{itemize}
\item[(a)] Draw $x_n'\sim q(\cdot | x_n^{(i-1)})$ and $u_{n-1}'\sim K(\cdot|u_{n-1}^{(i)})$. 
Compute $\hat{p}_{u_{n-1}'}^{(m)}(x'_n|x^{(i)}_{n-1},\theta^{(i)})$ by running iteration $n-1$ of Algorithm~\ref{imp}.
\item[(b)] With probability $\alpha(x_n',u_{n-1}'|x_n^{(i-1)},u_{n-1}^{(i)},x_{n-1}^{(i)},\theta^{(i)})$ given by (\ref{aprobx3}) 
put $x^{(i)}_n=x'_n$ and $u_{n-1}^{(i)}=u_{n-1}'$ otherwise store the current value $x^{(i)}_n=x^{(i-1)}_n$ (and $u_{n-1}^{(i)}$ remains unchanged). 
\end{itemize}
\item If $i=n_{\textrm{iters}}$, stop. Otherwise, set $i:=i+1$ and go to step 2.
\end{enumerate}

\textbf{Output:} $\theta^{(1)},\ldots,\theta^{(n_{\textrm{iters}})}$, $x^{o,(1)},\ldots,x^{o,(n_{\textrm{iters}})}$.
\end{algorithm}

{As presented, Algorithm~\ref{algaug} is appropriate for the general case of 
noisy and partial observation of $X_t$. In the special case of data consisting of noise 
free observation of all SDE components (so that $\Sigma=0$ and $F=I_d$ in (\ref{eqn:obs})), 
steps 3-5 are not required. Additionally, step 2 should propose the full auxiliary vector 
$u'$ from $K(u'|u)$ in (\ref{cn}) . Hence, this special case corresponds to the 
CPMMH algorithm with $\hat{p}_{u'}^{(m)}({y}|\theta')$ obtained by importance sampling 
and the acceptance probability is as in (\ref{aprob}). In the case of noise free 
observation of a subset of components of $X_t$, the scheme proceeds as in Algorithm~\ref{algaug}, with the unobserved components 
of $X_t$ updated in steps 3-5.}

\subsection{Connection with existing samplers for SDEs}
\label{sec:existing}
Consider aCPMMH with $N=1$ particle and $\rho=0$. In this case aCPMMH exactly coincides with the modified innovation 
scheme introduced by \cite{GoliWilk08} \citep[see also][]{GoliWilk10,Papaspiliopoulos13,Fuchs_2013,Meulen17}. We note that 
for this choice of $N$ there is a one-to-one correspondence between the innovations $u$ and the latent path $x$. Hence, step 2 of 
the Gibbs sampler in Section~\ref{sec:apmmhAlg} is equivalent to directly updating the latent path $x$ in blocks of size 
$2m-1$. To make this clear, consider updating $x_{(t-1,t+1)}$. Upon substituting (\ref{transEst}) into the acceptance probability in (\ref{aprobx1}) we obtain
\begin{align*}
& \min\left\{1\,,\,  \frac{p_{e}^{(m)}(x_{(t-1,t]}'|x_{t-1},\theta)}{p_{e}^{(m)}(x_{(t-1,t]}|x_{t-1},\theta)}\times 
 \frac{p_{e}^{(m)}(x_{(t,t+1]}'|x_t',\theta)}{p_{e}^{(m)}(x_{(t,t+1]}|x_t,\theta)}\right. \\
&\qquad\qquad\left. \times \frac{g(x_{(t-1,t)}|x_{t-1},x_{t},\theta)}{g(x_{(t-1,t)}'|x_{t-1},x_{t}',\theta)}\times \frac{g(x_{(t,t+1)}|x_t,x_{t+1},\theta)}{g(x_{(t,t+1)}'|x_t',x_{t+1},\theta)} \times\frac{p(y_t|x_t',\theta)}{p(y_t|x_t,\theta)} \times \frac{q(x_t | x_t')}{q(x_t' | x_t)}\right\}
\end{align*}
corresponding to a MH step that uses a RWM proposal to obtain $x_t'$ and then conditional on this value, uses the 
bridge construct to propose $x_{(t-1,t)}'$ and $x_{(t,t+1)}'$. Step 1 of the Gibbs sampler is equivalent to the reparameterisation 
used by the modified innovation scheme. Rather than update $\theta$ conditional on $x$ (and $y$), the innovations 
$u$ are the effective component being conditioned on. The motivation for this reparameterisation is to break down 
the well known problematic dependence between $\theta$ and $x$ \citep{StramRobe01}. To make the connection clear, note that upon 
combining (\ref{margll}) with (\ref{transEst}) and substituting the result into the acceptance probability in (\ref{aprob2}), 
we obtain
\begin{align*}
\min\left\{1\,,\,\frac{\pi(\theta')}{\pi(\theta)}\times \prod_{t=0}^{n-1} \frac{p_{e}^{(m)}(x_{(t,t+1]}|x_t,\theta')}
{p_{e}^{(m)}(x_{(t,t+1]}|x_t,\theta)}\times \prod_{t=0}^{n-1} \frac{g(x_{(t,t+1)}|x_t,x_{t+1},\theta)}{g(x_{(t,t+1)}|x_t,x_{t+1},\theta')}\times
\frac{p(y|x^o,\theta')}{p(y|x^o,\theta)}\times
\frac{q(\theta|\theta')}{q(\theta'|\theta)}\right\}.
\end{align*}
It is straightforward to show that the Jacobian associated 
with the change of variables (from $x$ to $u$) is given by $\prod_{t=0}^{n-1}g(x_{(t,t+1)}|x_t,x_{t+1},\theta)^{-1}$ and therefore the above acceptance probability coincides with that 
obtained for the parameter update in the modified innovation scheme \citep[see e.g. page 14 of][]{GoliWilk10}.

For $N=1$ and $0<\rho<1$, aCPMMH can be seen as an extension of the modified innovation scheme that uses a Crank-Nicolson proposal 
for the innovations. A recent application can be found in \cite{Arnaudon20}. We assess the performance of aCPMMH for 
different values of $\rho$ and $N$ in Section~\ref{sec:app}.

\subsection{Initialisation and tuning choices}
\label{sec:init}

Recall that both CPMMH and PMMH require setting the number of particles $N$ and, if using a random walk Metropolis (RWM) 
proposal, a suitable innovation variance. Practical advice on choosing $N$ for (C)PMMH is discussed at the end of 
Sections~\ref{sec:pmmh} and \ref{sec:cpmmh}. For a RWM proposal of the form $q(\theta'|\theta)=N(\theta^*;\theta,\Omega)$, 
a rule of thumb for the innovation variance $\Omega$ is to take $\Omega=\frac{2.56^2}{p}\widehat{\textrm{var}}(\theta|y)$ 
\citep{Sherlock2015}, which could be obtained from an initial pilot run (such as that required to find a plausible 
$\theta$ value for subsequently choosing $N$). 

For (C)PMMH, both the pilot and main monitoring runs require careful initialisation of $\theta$ \citep{Owen15}. 
The aCPMMH scheme additionally requires initialisation of $x^o$, with poor choices likely to slow initial convergence 
of the Gibbs sampler. One possibility is to seek an approximation to $\pi^{(m)}(\theta,x^o|y)$, 
denoted $\pi^{(a)}(\theta,x^o|y)$, for which samples can be obtained (e.g. via MCMC) at relatively low computational cost. These samples 
can then be used to compute estimates $\hat{\textrm{E}}(\theta|y)$ and $\hat{\textrm{E}}(x_t|y)$, which can be used to initialise aCPMMH. Further, the proposal variances for $\theta$ and $x_t$ can be made proportional to the estimates $\widehat{\textrm{var}}(\theta|y)$ and         $\widehat{\textrm{var}}(x_t|y)$ respectively, which can also be computed from the samples. For SDE models of the form (\ref{eqn:sdmem}), the linear noise approximation (LNA) \citep{stathopoulos13,fearnhead14} provides a tractable Gaussian  approximation. We describe the LNA, its solution and sampling of $\pi^{(a)}(\theta,x^o|y)$ in Appendix~\ref{lna}. In scenarios where using the LNA is not practical, we suggest initialising a pilot run of aCPMMH with $x^o=y$ (if $d_o=d$ so that all components are observed) or sampling $x^o$ via recursive application of the bridge construct in (\ref{bridge}). The pilot run can be used to obtain further quantities required for tuning the proposal densities $q(\theta'|\theta)$ and $q(x_t'|x_t)$. Hence, our intitialisation and tuning advice can be summarised by the following two options:
\begin{enumerate}
\item Perform a short pilot run of an MCMC scheme targeting $\pi^{(a)}(\theta,x^o|y)$ (as described in Appendix~\ref{lna} for the LNA) to obtain the estimates $\hat{\textrm{E}}(\theta|y)$, $\hat{\textrm{E}}(x_t|y)$, $\widehat{\textrm{var}}(\theta|y)$ and $\widehat{\textrm{var}}(x_t|y)$. These quantities are used to initialise the main monitoring run of aCPMMH and in the RWM proposals for $\theta$ and the components of $x^o$.
\item Perform a short pilot run of aCPMMH with $x^o$ initialised at the data $y$ (if all SDE components are observed) or, in the case of incomplete observation of all SDE components, recursively draw from (\ref{bridge}), retaining only those values at the observation times. Compute estimates as in option 1, for use in the main monitoring run. 
\end{enumerate}
The length of the pilot run can be set by choosing a fraction of the main monitoring run to fix the overall computational budget. For simplicity, we use RWM proposals in the pilot runs with diagonal innovation variances chosen to obtain (approximately) a desired acceptance rate. We note that Option 2 additionally requires specifying an initial number of particles $N$ for the pilot run. In our experiments, we find that $N=1$ is often sufficient. For either option, the number of particles can be further tuned if desired, before the main monitoring run.

\section{Applications}
\label{sec:app}

We consider three applications of increasing complexity. All algorithms are coded in R and were run on a desktop computer with an Intel quad-core CPU. 
For all experiments, we compare the performance of competing algorithms using minimum (over each parameter chain and for aCPMMH, $x^o$ chain) 
effective sample size per second (mESS/s). Effective sample size (ESS) is the number of independent and identically distributed samples from the target that would produce an estimator with the same variance as the auto-correlated MCMC output. We computed ESS using the R coda package, details of which can be found in \cite{Plummer06}. {When running CPMMH and aCPMMH with $\rho>0$ we set $\rho=0.99$. This pragmatic choice promotes good mixing in the $\theta$ chain (for CPMMH and aCPMMH) and $x^o$ chain (for aCPMMH) while allowing the auxiliary variables to mix on a scale comparable to $\theta$ and $x^o$.} We report CPU time based on the main monitoring runs and note that CPU cost of tuning was small relative to the cost of the main run (and typically less than $10\%$ of the reported CPU time). For all experiments (unless stated otherwise) we used a discretisation of $\Delta\tau=0.2$ which we found gave a good balance between accuracy (in the sense of limiting discretisation bias) and computational performance.   

\subsection{Square-root diffusion process}
\label{sec.BDmodel}

\begin{figure*}[t]
\centering
\psfragscanon
\psfrag{yt}[][][0.7][-90]{$Y_t$}
\psfrag{t}[][][0.7][0]{$t$}
\includegraphics[width=7cm,height=17cm,angle=-90]{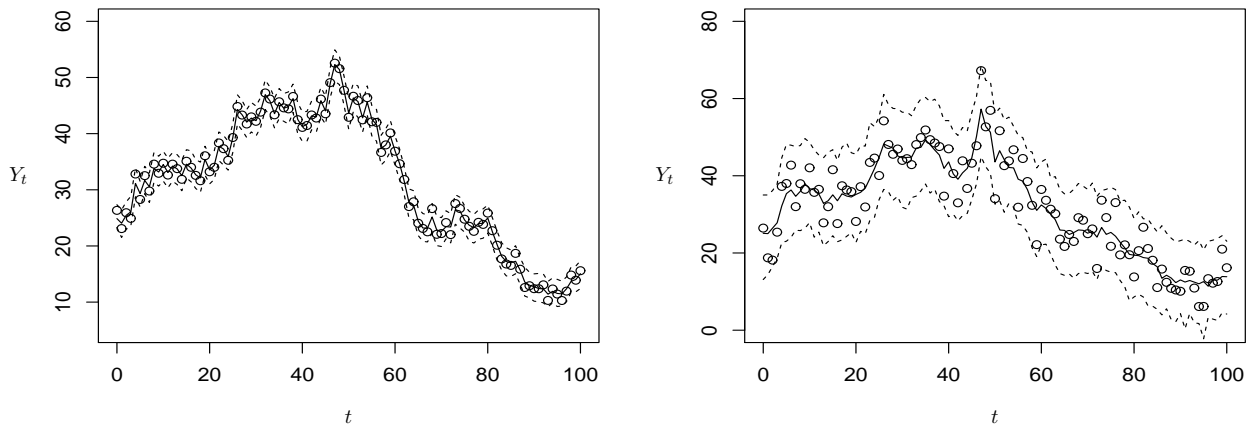}
\caption{Birth--Death model. Data sets (circles) and summaries (mean and $95\%$ credible intervals obtained from 
the output of aCPMMH) of the within-sample predictive $\pi(y|\mathcal{D}_1)$ (left) and $\pi(y|\mathcal{D}_2)$ (right).}
\label{fig:figBD2}
\end{figure*}

Consider a univariate diffusion process satisfying an It\^o SDE of the form
\begin{equation}\label{feller}
dX_{t}=\left(\theta_{1}-\theta_{2}\right)X_{t}\,dt + \sqrt{\left(\theta_{1}+\theta_{2}\right)X_{t}}\,dW_{t},
\end{equation}
which can be seen as a degenerate case of a Feller square-root diffusion \citep{feller52}. 
We generated two synthetic data sets consisting of 101 observations 
at integer times using $\theta=(0.05,0.06)^T$ and a known initial 
condition of $X_{0}=25$. The observation model is $Y_t\sim N(X_t,\sigma^2)$ where 
$\sigma\in\{1,5\}$ giving data sets designated as $\mathcal{D}_1$ and $\mathcal{D}_2$  
respectively (and shown in Figure~\ref{fig:figBD2}). 
We took independent $N(0,10^2)$ priors for each $\log \theta_i$, $i=1,2$, and work 
on the logarithmic scale when using the random walk proposal mechanism.

We ran aCPMMH for $50K$ iterations with $\rho$ fixed at $0.99$. We report results for 
$N=1$ particle, since $N>1$ gave no increase in overall performance. Both tuning 
and initialisation methods (options 1 and 2 of Section~\ref{sec:init}) were implemented 
and denoted ``aCPMMH (1)" and ``aCPMMH (2)". 
We additionally include results based on the output of PMMH and CPMMH, which were tuned in line with the guidance 
given at the end of Sections~\ref{sec:pmmh} and \ref{sec:cpmmh}.

\begin{figure*}[t]
\centering
\psfragscanon
\psfrag{thet1}[][][0.7][0]{$\theta_1$}
\psfrag{thet2}[][][0.7][0]{$\theta_2$}
\psfrag{thetDiff}[][][0.7][0]{$\theta_1 - \theta_2$}
\includegraphics[width=5.7cm,height=17cm,angle=-90]{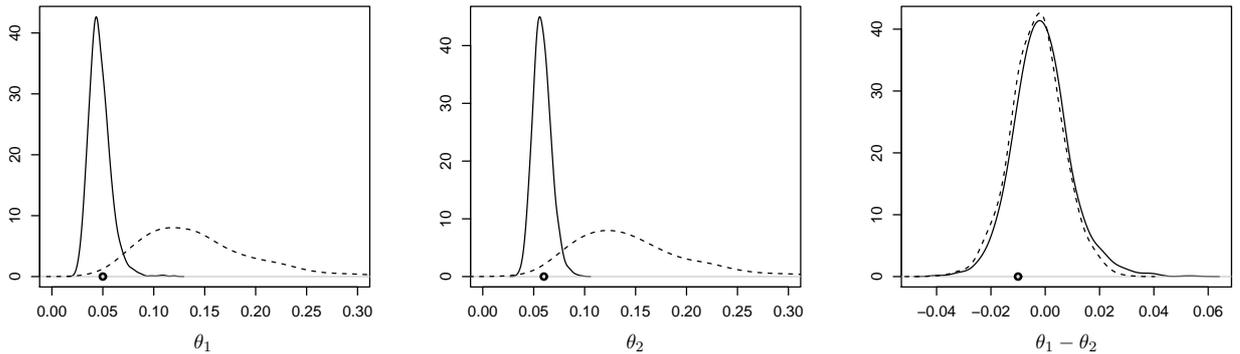}
\caption{Birth--Death model. Marginal posterior distributions using data set $\mathcal{D}_2$ and based on the output of
aCPMMH (solid lines) and the LNA (dashed lines). The true values of $\theta_1$, $\theta_2$ and $\theta_1 - \theta_2$ are indicated.}
\label{fig:figBD}
\end{figure*}

Table~\ref{tab:tabBD} and Figure~\ref{fig:figBD} summarise our results. The latter shows marginal 
posteriors obtained from the output of aCPMMH, and for comparison, from the LNA as an inferential model. 
We note substantial differences in posteriors obtained when using the discretised SDE in (\ref{feller}) as the inferential 
model compared to inferences made under the LNA. This is not surprising since for this example, the ground 
truth $\theta_1$ and $\theta_2$ values are similar, for which the assumption that fluctutaions about the 
mean of $X_t$ are small (as is necessary for an accurate LNA), is unreasonable. Nevertheless, we were still 
able to use the LNA to adequately initialise and tune aCPMMH. We also note that both initialisation 
and tuning options give comparable results.

It is evident from Table~\ref{tab:tabBD} that aCPMMH offers substantial improvements in overall efficiency 
compared to PMMH and, to a lesser extent, CPMMH. Minimum effective sample size per second for PMMH : CPMMH : aCPMMH 
scales as $1:2.7:6$ for data set $\mathcal{D}_{1}$ and $1:2.5:2.5$ for data set $\mathcal{D}_{2}$. We found that 
as the measurement error variance ($\sigma^2$) is increased, the optimal number of particles $N$ for both PMMH and CPMMH 
also increased. Although aCPMMH required $N=1$ (that is, we observed no additional improvement in overall efficiency 
for $N>1$), the mixing deteriorates, due to having to integrate over the additional uncertainty in the latent process 
at the observation times. Finally, although the Euclidean sorting algorithm used in CPMMH is likely to be effective 
for this simple univariate example, we anticipate its deterioration in subsequent examples with increasing state 
dimension.     

\begin{table*}[t]
  \centering
  \small
  \begin{tabular}{@{}llllllll@{}}
    \toprule
    Data set & Algorithm  & $\rho$ & $N$ & CPU(s) & mESS $(\theta,x^o)$ & mESS/s & Rel.  \\
    \midrule
            &             &           &             &                &      &       &  \\    
$\mathcal{D}_{1}$ ($\sigma=1$)  &aCPMMH (1)    &0.99 &\phantom{0}1 &\phantom{0}777&(3194, 3818)  &4.11  &6.0 \\
                                &aCPMMH (2)    &0.99 &\phantom{0}1 &\phantom{0}780&(3080, 3757)  &3.95  &5.8 \\
                     & \phantom{a}CPMMH        &0.99 &\phantom{0}2 &\phantom{0}745&(1358, \phantom{0}--\phantom{0})  &1.82  &2.7 \\
                    & \phantom{aC}PMMH         &0.00&10                     &2217&(1513, \phantom{0}--\phantom{0})  &0.68  &1.0 \\
          &             &           &             &                &      &       &  \\ 
$\mathcal{D}_{2}$ ($\sigma=5$)  &aCPMMH (1)    &0.99 &\phantom{0}1 &\phantom{0}775&(829, 481)  &0.62  &2.5 \\
                                &aCPMMH (2)    &0.99 &\phantom{0}1 &\phantom{0}770&(799, 450)  &0.58  &2.3 \\
                     & \phantom{a}CPMMH        &0.99 &\phantom{0}6            &1684&(1050, \phantom{0}--\phantom{0}) &0.62  &2.5 \\
                    & \phantom{aC}PMMH         &0.00 &20                      &4989&(1254, \phantom{0}--\phantom{0}) &0.25  &1.0 \\
    \bottomrule
  \end{tabular}
  \caption{Birth--Death model. Number of particles $N$, correlation parameter $\rho$, CPU time (in seconds $s$), minimum ESS (over $\theta$ and $x^o$ chains), minimum ESS per second and relative (to PMMH) minimum ESS per second. All results are based on $5\times 10^4$ iterations of each scheme.}\label{tab:tabBD}	
\end{table*}

\subsection{Lotka-Volterra}
\label{sec.LVmodel}

The Lotka-Volterra system describes the time-course behaviour of two interacting 
species: prey $X_{1,t}$ and predators $X_{2,t}$. The stochastic differential equation 
describing the dynamics of $X_t=(X_{1,t},X_{2,t})^T$ is given by
\begin{align}
& d\begin{pmatrix}X_{1}\\ X_{2}\end{pmatrix}= \begin{pmatrix}
				\theta_{1}X_{1}-\theta_{2}X_{1}X_{2} \\
				\theta_{2}X_{1}X_{2}-\theta_{3}X_{2}
			\end{pmatrix}\,dt + \begin{pmatrix}\theta_{1}X_{1}+\theta_{2}X_{1}X_{2} & -\theta_{2}X_{1}X_{2} \\
			 -\theta_{2}X_{1}X_{2}	 & \theta_{2}X_{1}X_{2}+\theta_{3}X_{2}
			\end{pmatrix}^{\frac{1}{2}}\,d \begin{pmatrix}W_{1}\\ W_{2}\end{pmatrix} \label{lv}
\end{align}
after suppressing dependence on $t$. 

\begin{figure*}[t]
\centering
\psfragscanon
\psfrag{y1t}[][][0.7][-90]{$Y_{1,t}$}
\psfrag{y2t}[][][0.7][-90]{$Y_{2,t}$}
\psfrag{t}[][][0.7][0]{$t$}
\includegraphics[width=7cm,height=17cm,angle=-90]{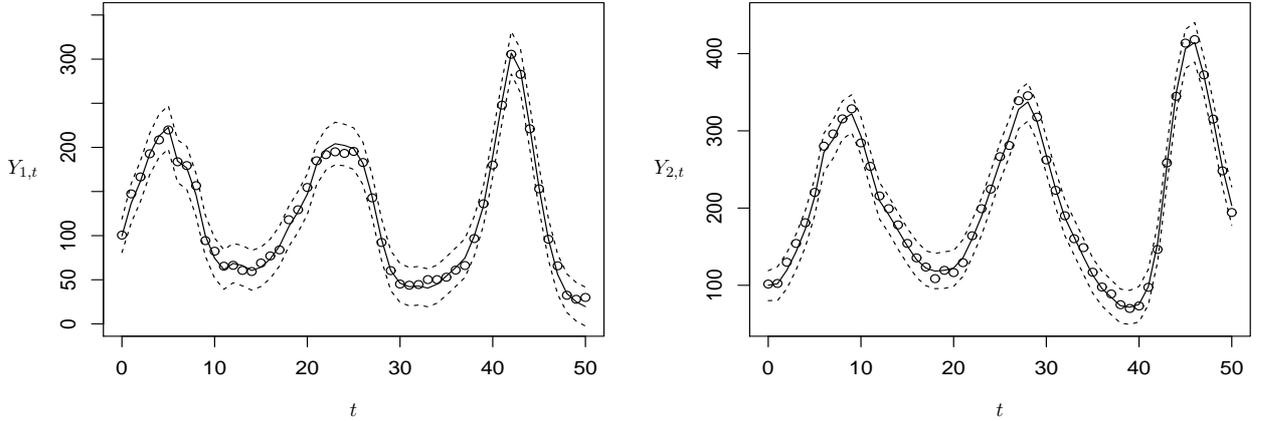}
\caption{Lotka--Volterra model. Data set $\mathcal{D}_3$ (circles) and summaries (mean and $95\%$ credible intervals obtained from 
the output of aCPMMH) of the within-sample predictive $\pi(y|\mathcal{D}_3)$ (left: prey component, right: predator component).}
\label{fig:figLV}
\end{figure*}

We repeated the experiments of \cite{GoliBrad19} which, for this example, involved 
three synthetic data sets generated with $\theta=(0.5,0.0025,0.3)^T$ and a known initial 
condition of $X_{0}=(100,100)^T$. The observation model is
\[
Y_{t}\sim \textrm{N}\left(X_t,\sigma^{2}I_{2}\right)
\]
where $I_{2}$ is the $2\times 2$ identity matrix and $\sigma\in\{1,5,10\}$ 
giving data sets designated as $\mathcal{D}_1$, $\mathcal{D}_2$ and $\mathcal{D}_3$ 
respectively. Data set $\mathcal{D}_3$ is shown in Figure~\ref{fig:figLV}, and gives dynamics typical 
of the parameter choice taken. The parameters correspond to the rates of prey reproduction, 
prey death and predator reproduction, and predator death. As the parameters must be 
strictly positive, we work on the logarithmic scale with independent $N(0,10^2)$ priors 
assumed for each $\log \theta_i$, $i=1,2,3$. The main monitoring runs consisted of 
$10^5$ iterations of aCPMMH, CPMMH (with $\rho=0.99$) and PMMH. Note that aCPMMH 
used random walk proposals in the $\log\theta$ and $x^o$ updates, with variances obtained from the output of 
an MCMC pilot run based on the LNA, which was also used to initialise $\theta$ and $x^o$. 

\begin{figure*}[t]
\centering
\psfragscanon
\psfrag{thet1}[][][0.7][0]{$\theta_1$}
\psfrag{thet2}[][][0.7][0]{$\theta_2$}
\psfrag{thet3}[][][0.7][0]{$\theta_3$}
\includegraphics[width=5.7cm,height=17cm,angle=-90]{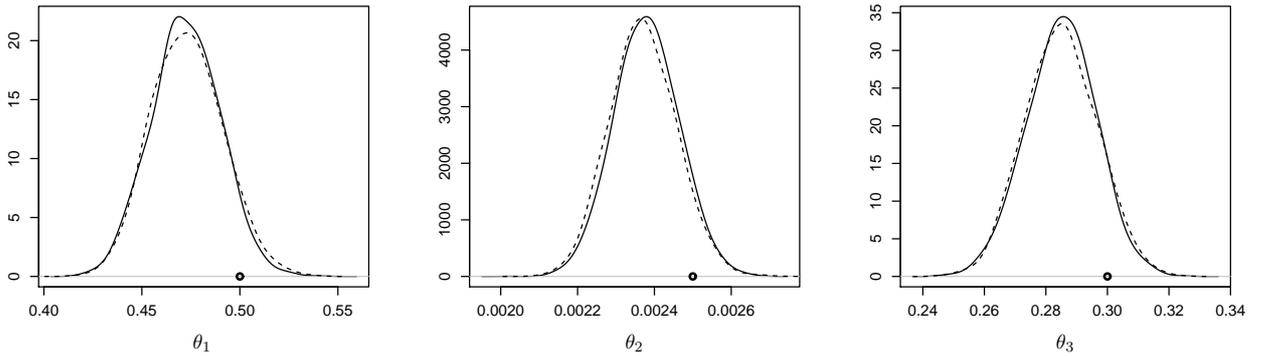}
\caption{Lotka--Volterra model. Marginal posterior distributions using data set $\mathcal{D}_3$ and based on the output of
aCPMMH (solid lines) and the LNA (dashed lines). The true values of $\theta_1$, $\theta_2$ and $\theta_3$ are indicated.}
\label{fig:figLV2}
\end{figure*}

From Figure~\ref{fig:figLV2} we see that aCPMMH gives parameter posterior output that 
is consistent with the ground truth (and also with the output of PMMH and CPMMH -- not shown). 
In this case, the LNA gives accurate output when used as an inferential model. We compare 
efficiency of PMMH, CPMMH and aCPMMH in Table~\ref{tab:tabLV}. We found that $N=1$ was sufficient 
for aCPMMH but also include results for $N=2$, which gave a small increase in minimum ESS but a decrease 
in overall efficiency, due to the increase (doubling) in CPU time. It is clear that as $\sigma$ increases, 
PMMH and CPMMH require an increase in $N$ to maintain a reasonable minimum ESS. Consequently, their performance 
degrades. Although the statistical efficiency (mESS) of aCPMMH reduces as $\sigma$ increases, the reduction 
is gradual (compared to that of CPMMH) and we see an increase in overall efficiency of aCPMMH (with $\rho=0.99$) 
of an order of magnitude over PMMH in all experiments, and over CPMMH for data sets $\mathcal{D}_2$ 
and $\mathcal{D}_3$. We also include the output of aCPMMH with $N=1$ and $\rho=0.0$, corresponding to the 
modified innovation scheme of \cite{GoliWilk08} (and as discussed in Section~\ref{sec:existing}). Although 
this approach works well compared to CPMMH and PMMH, and gives well mixing parameter chains, we see a decrease 
in mESS (relative to aCPMMH with $\rho=0.99$) calculated from the $x^o$ chains, and this relative difference increases 
as $\sigma$ increases.    

\begin{figure*}[t]
\centering
\psfragscanon
\psfrag{log10Dt}[][][0.7][0]{$\log_{10}\Delta \tau$}
\psfrag{log2CPUdiff}[][][0.7][0]{$\Delta\log_{2}\textrm{CPU}$}
\includegraphics[width=5.7cm,height=17cm,angle=-90]{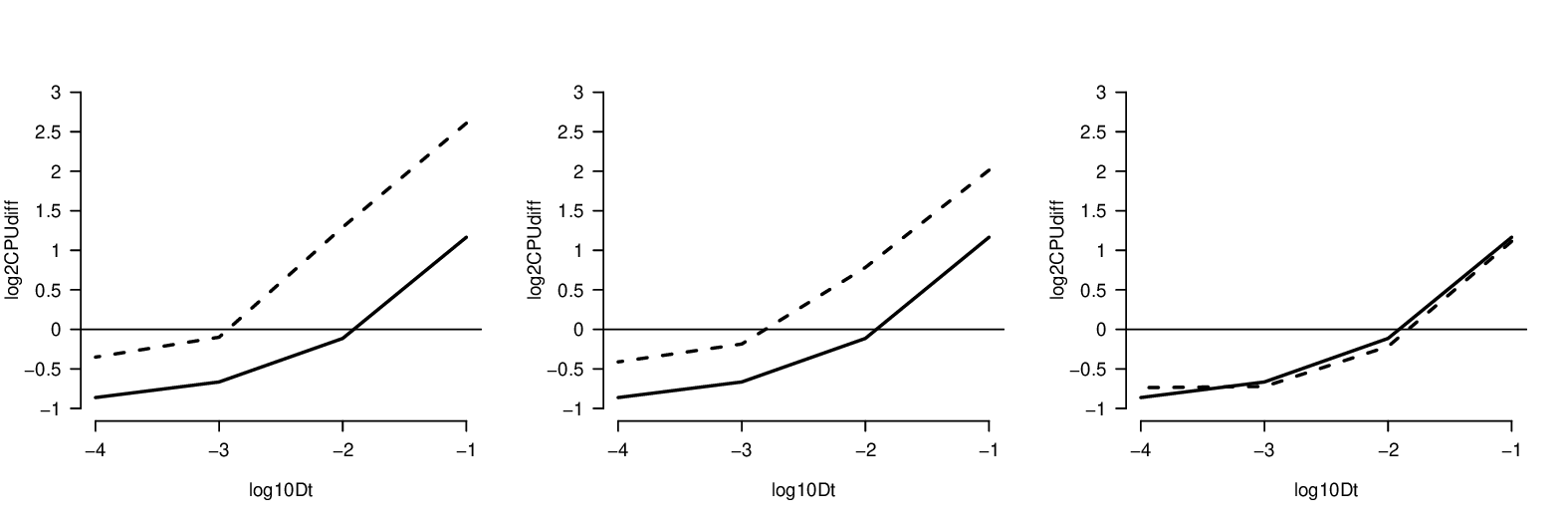}
\caption{Lotka--Volterra model. Difference (2 cores vs 1 core) in $\log_2$ CPU times ($\Delta\log_{2}\textrm{CPU}$) against 
$\log_{10} \Delta \tau$ using aCPMMH with $N=1$ (solid lines) and CPMMH (dashed lines) with $N=3$ (data set $\mathcal{D}_1$, left), $N=8$ (data set $\mathcal{D}_2$, centre) and $N=18$ (data set $\mathcal{D}_3$, right).}
\label{fig:figLV3}
\end{figure*}

Finally, we compare the performance of CPMMH and aCPMMH when parallelised over two cores. For aCPMMH 
(and as discussed at the end of Section~\ref{sec:apmmhAlg}), operations over $t$ in steps 2,3 and 4 
of Algorithm~\ref{algaug} can be performed in parallel. For CPMMH, we perform the propagate step of the 
particle filter (step 2(b) of Algorithm~\ref{PF}) in parallel. Figure~\ref{fig:figLV3} shows 
the difference (2 cores vs 1 core) in $\log_2$ CPU times (denoted $\Delta\log_{2}\textrm{CPU}$) against 
$\log_{10} \Delta \tau$, where the discretisation level is $\Delta\tau\in\{10^{-4},10^{-3},10^{-2},10^{-1}\}$; 
for a perfect speed up from the use of two cores, this would be $-1$. 
Results based on aCPMMH used $N=1$ in all cases whereas CPMMH used $N=3$, $N=8$ and $N=18$. These values correspond 
to the the numbers of particles required for each synthetic data set. For aCPMMH, we see that using 2 cores 
is beneficial for $\Delta\tau \leq 10^{-2}$. For CPMMH and $N=1$, there is almost no benefit in a multi-core 
approach (and CPU time using 2 cores is typically higher than a single core approach). This is unsurprising 
given the resampling steps (performed in serial) between the propagate steps. As $N$ increases, the benefit 
of using 2 cores can be seen.

\begin{table*}[t]
  \centering
  \small
  \begin{tabular}{@{}llllllll@{}}
    \toprule
    Data set & Algorithm  & $\rho$ & $N$ & CPU(s) & mESS $(\theta,x^o)$ & mESS/s & Rel.  \\
    \midrule
            &             &           &             &                &      &       &  \\    
$\mathcal{D}_{1}$ ($\sigma=1$)  &aCPMMH&0.99 &\phantom{0}1 &\phantom{00}7330&(9375, 9483)  &1.279  &27.8 \\
				&      &0.99 &\phantom{0}2 &\phantom{0}12773&(9446, 12485) &0.740  &16.1 \\
				&      &0.00 &\phantom{0}1 &\phantom{00}6877&(8805, 2028)  &0.299  &\phantom{0}6.5 \\
                     & \phantom{a}CPMMH&0.99 &\phantom{0}3 &\phantom{0}11280&(8023, \phantom{0}--\phantom{0})  &0.711  &16.3 \\
                     & \phantom{aC}PMMH&0.00 &16           &\phantom{0}59730&(2771, \phantom{0}--\phantom{0})  &0.046  &\phantom{0}1.0 \\
          &             &           &             &                &      &       &  \\ 
$\mathcal{D}_{2}$ ($\sigma=5$)  &aCPMMH&0.99 &\phantom{0}1 &\phantom{00}6780&(7331, 6807)  &1.004  &25.7 \\
				&      &0.99 &\phantom{0}2 &\phantom{0}12807&(7877, 7117)  &0.556  &14.2 \\
				&      &0.00 &\phantom{0}1 &\phantom{00}6769&(8022, 1380)  &0.204  &\phantom{0}5.2 \\
                     & \phantom{a}CPMMH&0.99 &\phantom{0}8 &\phantom{0}29780&(3681, \phantom{0}--\phantom{0}) &0.124  &\phantom{0}3.2 \\
                    & \phantom{aC}PMMH &0.00 &20           &\phantom{0}75930&(2959, \phantom{0}--\phantom{0}) &0.039  &\phantom{0}1.0 \\
           &             &           &             &                &      &       &  \\   
$\mathcal{D}_{3}$ ($\sigma=10$) &aCPMMH&0.99 &\phantom{0}1 &\phantom{00}6772 &(4986, 3301)  &0.487  &16.8 \\
				&      &0.99 &\phantom{0}2 &\phantom{0}12753&(5859, 3446)   &0.270  &\phantom{0}9.3 \\
				&      &0.00 &\phantom{0}1 &\phantom{00}6786&(4676, 1384)   &0.203  &\phantom{0}7.0 \\
                     &\phantom{a}CPMMH &0.99 &19           &\phantom{0}71520 &(3516, \phantom{0}--\phantom{0})  &0.049  &\phantom{0}1.7 \\
                    &\phantom{aC}PMMH  &0.00 &28           &105770           &(3031, \phantom{0}--\phantom{0})  &0.029  &\phantom{0}1.0 \\
    \bottomrule
  \end{tabular}
  \caption{Lotka--Volterra model. Number of particles $N$, correlation parameter $\rho$, CPU time (in seconds $s$), minimum ESS (over $\theta$ and $x^o$ chains), minimum ESS per second and relative (to PMMH) minimum ESS per second. All results are based on $10^5$ iterations of each scheme.}\label{tab:tabLV}	
\end{table*}

\subsection{Autoregulatory gene network}
\label{sec.ARmodel}

A commonly used mechanism for
auto-regulation in prokaryotes which has been well-studied and
modelled is a negative feedback mechanism whereby dimers of a protein
repress its own transcription \citep[e.g.][]{ARM98}. A 
simplified model for such a prokaryotic auto-regulation, 
based on this mechanism can be 
found in \cite{GoliWilk05} \cite[see also][]{GoliWilk11}. 
We consider the SDE representation of the dynamics of 
the key species involved in this mechanism. These are 
$\textsf{RNA}$, $\textsf{P}$, $\textsf{P}_2$ and $\textsf{DNA}$, 
denoted as $X_1$, $X_2$, $X_3$ and $X_4$ respectively. 
The SDE takes the form (\ref{eqn:sdmem}) where
\[
\alpha(X_t,\theta)=S\,h(X_t,\theta),\qquad \beta(X_t,\theta)= S\,h(X_t,\theta)\, S^T,
\]
the stoichiometry matrix $S$ is
\[
S = \left(\begin{array}{rrrrrrrr}
0&\phantom{-}0&\phantom{-}1&\phantom{-}0&0&0&-1&0\\
0&0&0&1&-2&2&0&-1\\
-1&1&0&0&1&-1&0&0\\
-1&1&0&0&0&0&0&0\\
\end{array}\right),
\]
and the hazard function $h(X_t,\theta)$ is
\begin{align*}
 h(X,\theta) &= (0.1 X_{4}X_{3}, \theta_1(10-X_{4}),
\theta_2 X_{4}, 0.2 X_{1}, 0.1 X_{2}(X_{2}-1)/2,
\theta_3 X_{3}, \theta_4 X_{1}, \theta_8 X_{2} )^T
\end{align*}
after dropping $t$ to ease the notation. Further details regarding the derivation 
of the SDE can be found in \cite{GoliWilk05}.

\begin{figure*}[t]
\centering
\psfragscanon
\psfrag{y1t}[][][0.7][-90]{$Y_{1,t}$}
\psfrag{y2t}[][][0.7][-90]{$Y_{2,t}$}
\psfrag{y3t}[][][0.7][-90]{$Y_{3,t}$}
\psfrag{y4t}[][][0.7][-90]{$Y_{4,t}$}
\psfrag{t}[][][0.7][0]{$t$}
\includegraphics[width=7cm,height=17cm,angle=-90]{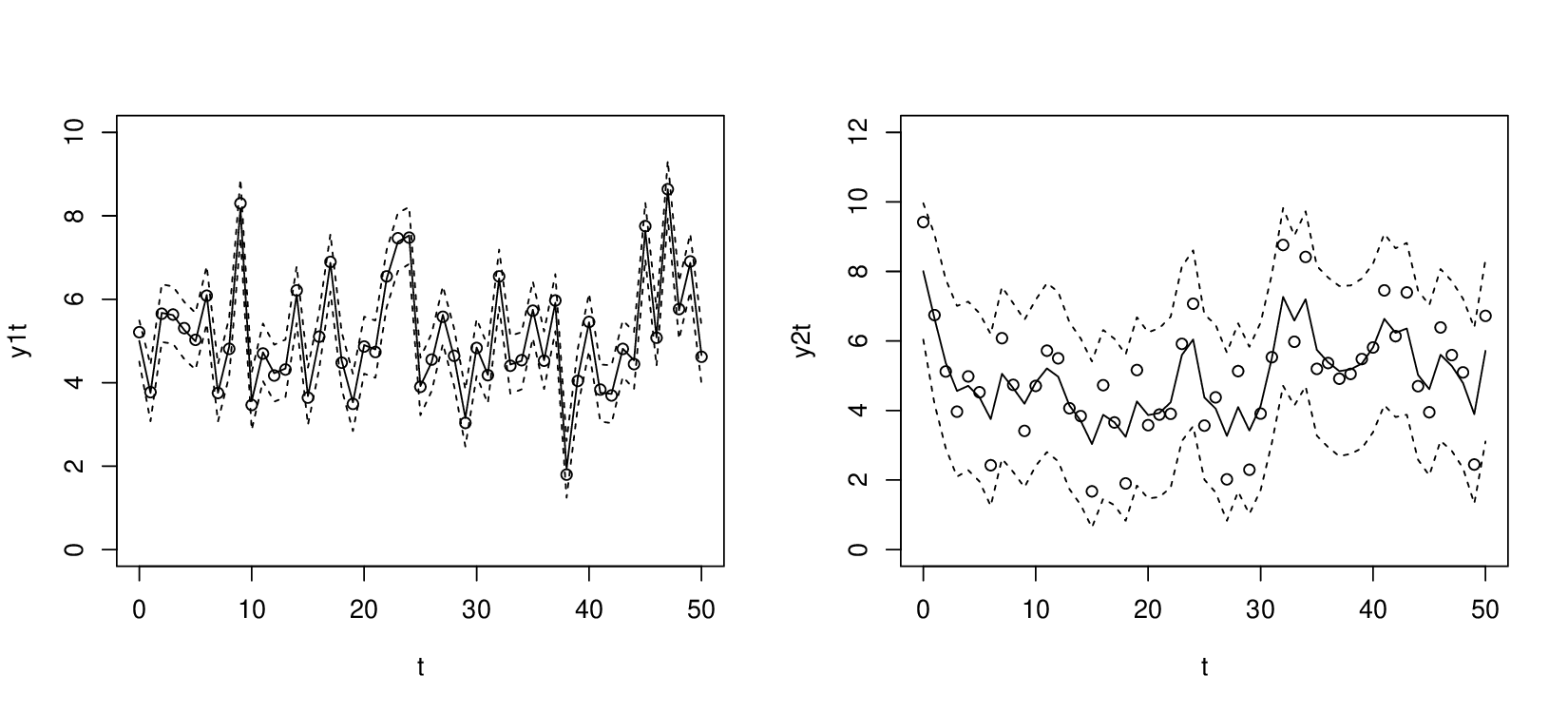}
\includegraphics[width=7cm,height=17cm,angle=-90]{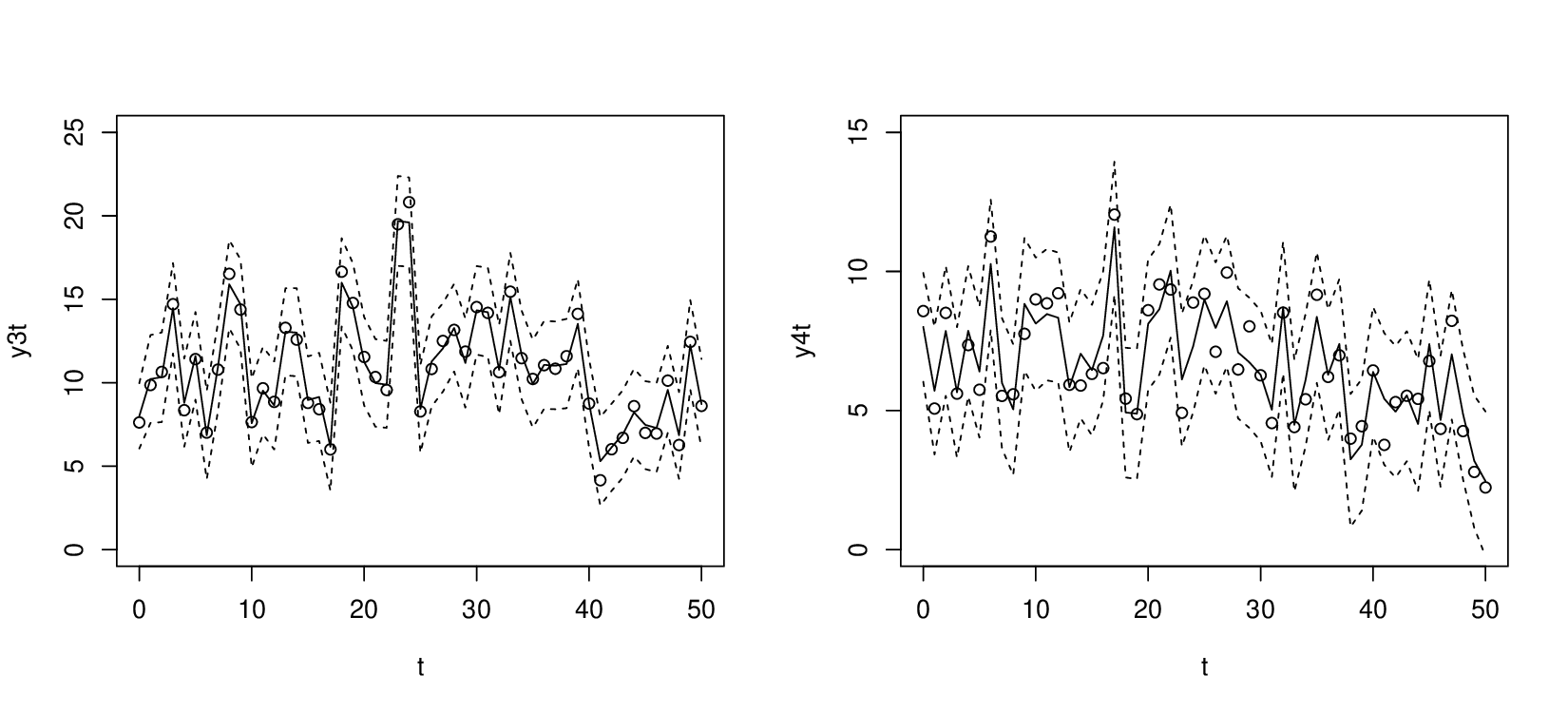}
\caption{Autoregulatory model. Data set (circles) and summaries (mean and $95\%$ credible intervals obtained from 
the output of aCPMMH) of the within-sample predictive $\pi(y|\mathcal{D})$.}
\label{fig:figAR}
\end{figure*}

The parameters $\theta=(\theta_{1},\theta_{2},\theta_{3},\theta_{4})^T$ correspond to the 
rate of protein unbinding at an operator site, the rate of transcription of a gene into mRNA, 
the rate at which protein dimers disassociate and the rate at which protein molecules degrade. 
We generated a single synthetic data set with $\theta=(0.7,0.35,0.9,0.3)^T$ and an initial 
condition of $X_{0}=(8,8,8,5)^T$. The observation model is
\[
Y_{t}\sim \textrm{N}\left(X_t,\Sigma\right)
\]
where $\Sigma$ is a diagonal matrix with elements $1,1,1,0.25$. The data are shown in Figure~\ref{fig:figAR}. 
Independent $U(-5,5)$ priors were assumed for each $\log \theta_i$, $i=1,2,3,4$. A short MH run was 
performed using the LNA, to obtain estimates of $\textrm{var}(\log\theta |y)$ and $\textrm{var}(x_t|y)$
(to be used the innovation variances of the random walk proposal mechanisms in (a)CPMMH) and plausible values 
of $\theta$ and $x^o$ (to be used to initialise the main monitoring runs of (a)CPMMH). Pilot runs of 
aCPMMH and CPMMH suggested taking $N=1$ and $N=20$ for each respective scheme. We then ran 
aCPMMH and CPMMH for $10^5$ iterations with these tuning choices. Table~\ref{tab:tabAR} 
and Figure~\ref{fig:figAR} summarise our findings.

It is clear that aCPMMH with $\rho=0.99$ results in a considerable improvement in statistical efficiency 
over aCPMMH with $\rho=0.0$ (which is the modified innovation scheme). In particular, minimum ESS (calculated 
over the $x^o$ chains) is almost an order of magnitude higher for $\rho=0.99$ (866 vs 5524). An improvement 
in overall efficiency of aCPMMH over CPMMH is evident, irrespective of the choice of $\rho$. Increasing $N$ 
to $2$ gives results in better mixing of the $x^o$ chains, but no appreciable increase in minimum ESS over 
all chains.

\begin{table*}[t]
  \centering
  \small
  \begin{tabular}{@{}lllllll@{}}
    \toprule
     Algorithm  & $\rho$ & $N$ & CPU(s) & mESS $(\theta,x^o)$ & mESS/s & Rel.  \\
    \midrule
                         &           &             &                &      &       &  \\    
aCPMMH           &0.99 &\phantom{0}1 &\phantom{0}18248 &(992, 5524)                      &0.054  &13.5 \\
aCPMMH           &0.99 &\phantom{0}2 &\phantom{0}41578 &(766, 6210)                      &0.018  &\phantom{0}4.6 \\
aCPMMH           &0.00 &\phantom{0}1 &\phantom{0}18252 &(1358, 866)                      &0.028  &\phantom{0}6.9 \\
\phantom{a}CPMMH &0.99 &20           &199782           &(805, \phantom{0}--\phantom{0})  &0.004  &\phantom{0}1.0 \\
    \bottomrule
  \end{tabular}
  \caption{Autoregulatory model. Number of particles $N$, correlation parameter $\rho$, CPU time (in seconds $s$), minimum ESS (over $\theta$ and $x^o$ chains), minimum ESS per second and relative (to PMMH) minimum ESS per second. All results are based on $10^5$ iterations of each scheme.}\label{tab:tabAR}	
\end{table*}

\begin{figure*}[t]
\centering
\psfragscanon
\psfrag{thet1}[][][0.7][0]{$\theta_1$}
\psfrag{thet2}[][][0.7][0]{$\theta_2$}
\psfrag{thet3}[][][0.7][0]{$\theta_3$}
\psfrag{thet4}[][][0.7][0]{$\theta_4$}	
\includegraphics[width=5.7cm,height=17cm,angle=-90]{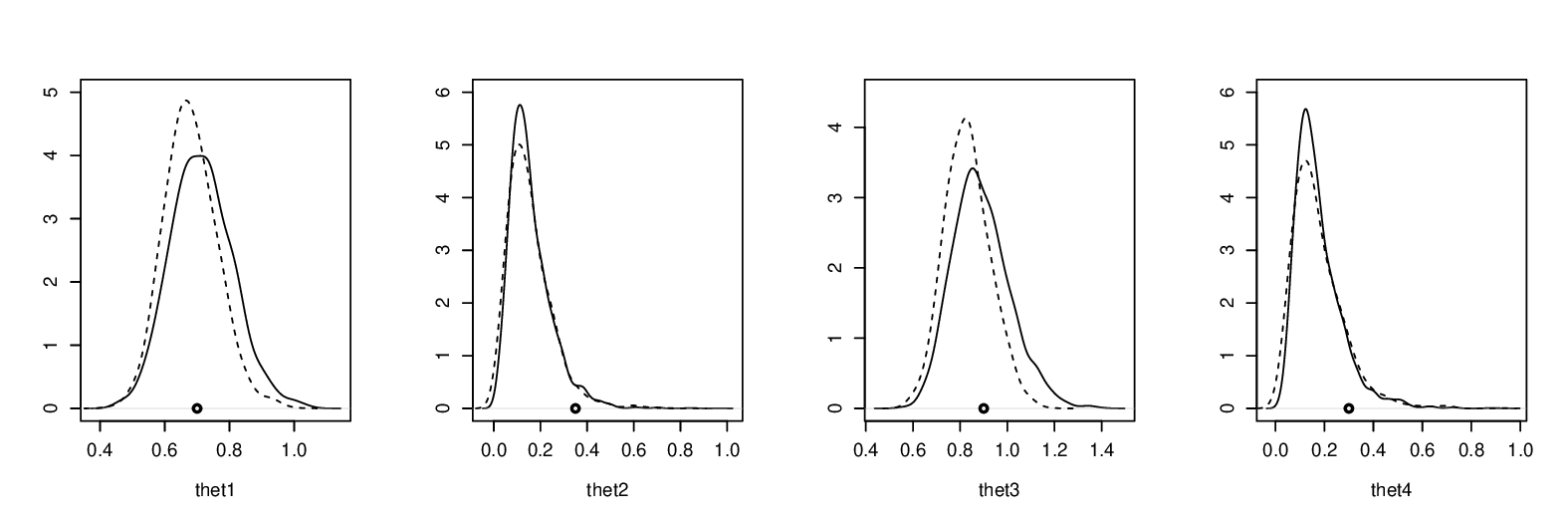}
\caption{Autoregulatory model. Marginal posterior distributions based on the output of
aCPMMH (solid lines) and the LNA (dashed lines). The true parameter values are indicated.}
\label{fig:figAR2}
\end{figure*}

\section{Discussion}
\label{sec:disc}

Given observations at discrete times, performing fully Bayesian 
inference for the parameters governing nonlinear, multivariate 
stochastic differential equations is a challenging problem. A discretisation approach allows 
inference for a wide class of SDE models, at the cost of introducing an additional bias 
(the so called discretisation bias). The simplest such approach uses the Euler-Maruyama 
approximation in combination with intermediate time points between observations, to allow 
a time step chosen by the practitioner, that should trade off computational cost and 
accuracy. It is worth emphasising that although a Gaussian transition density is assumed, 
the mean and variance are typically nonlinear functions of the diffusion process, and consequently, 
the data likelihood (after integrating out at intermediate times) remains intractable, even under the assumption of additive 
Gaussian noise.

These remarks motivate the use of pseudo-marginal Metropolis-Hastings (PMMH) schemes, which replace 
an evaluation of the intractable likelihood with a realisation of an unbiased estimator, obtained from a 
single run of a particle filter over dynamic states \citep{andrieu10}. It is crucial that the number of particles is carefully 
chosen to balance computational efficiency whilst allowing for reasonably accurate likelihood estimates. 
Inducing strong and positive correlation between successive likelihood estimates can reduce the variance 
of the acceptance ratio, permitting fewer particles \citep{dahlin2015,deligiannidis2018}. Essentially, 
the (assumed Gaussian) innovations that are used to construct the likelihood estimates are updated with 
a Crank-Nicolson (CN) proposal. The resampling steps in the particle filter are also modified in order to 
preserve correlation; the random numbers used during this step are included in the CN update, and the particle 
trajectories are sorted before resampling takes place at the next time point. We follow \cite{choppala2016} and 
\cite{GoliBrad19} by using a simple Euclidean sorting procedure based on the state of the particle trajectory 
at the current observation time. We find that the effectiveness of this correlated PMMH (CPMMH) 
approach degrades as the observation variance and state dimension increases.        
  
Our novel approach avoids the use of resampling steps, by updating parameters conditional on the values 
of the latent diffusion process at the observation times (and the observations themselves), whilst integrating 
over the state uncertainty at the intermediate times. An additional step is then used to update the 
latent process at the observation times, conditional on the parameters and data. The resulting algorithm can 
be seen as a pseudo-marginal scheme, with unbiased estimators of the likelihood terms obtained via importance 
sampling. We further block together the updating of the latent states and the innovations used to construct the 
likelihood estimates, and adopt a CN proposal mechanism for the latter. We denote the resulting sampler as 
augmented, correlated PMMH (aCPMMH). A related approach is given by \cite{fearnhead16}, who use particle 
MCMC with additional latent variables, carefully chosen to trade-off the error
in the particle filter against the mixing of the MCMC steps. We emphasise that unlike this approach, 
the motivation behind aCPMMH is to avoid use of a particle filter altogether, the benefits of which are two fold: 
positive correlation between successive likelihood estimates is preserved and the method for obtaining 
these likelihood estimates can be easily parallelised over observation intervals. Section~\ref{sec.LVmodel} shows 
that once the updating over an inter-observation interval is sufficiently costly substantial gains can 
be obtained by parallelising this task over the different inter-observation intervals: this could be 
useful for stiff SDEs, high-dimensional SDEs, or if multiple inter-observation intervals are 
tackled by a single importance sampler.

In addition to the tuning choices required by CPMMH (that is, the number of 
particles $N$, correlation parameter $\rho$ in the CN proposal and parameter proposal mechanism), aCPMMH 
requires initialisation of the latent process at the observation times and a suitable proposal mechanism. If a 
computationally cheap approximation of the joint posterior can be found, this may be used to initialise and 
tune aCPMMH. To this end, we found that use of a linear noise approximation (LNA) can work well, even in 
settings when inferences made under the LNA are noticeably discrepant, compared to those obtained under the SDE. 
In scenarios where use of the LNA is not practical, a pilot run of aCPMMH can be used instead.   

We compared the performance of aCPMMH with both PMMH and CPMMH using three examples of increasing complexity. 
In terms of overall efficiency (as measured by minimum effective sample size per second), aCPMMH offered 
an increase of up to a factor of 28 over PMMH. We obtained comparable performance with CPMMH for a 
univariate SDE application, and an increase of up to factors of 10 and 14 in two applications involving 
SDEs of dimension 2 and 4 respectively. Our experiments suggest that although the mixing efficiency of 
aCPMMH increases with $N$, the additional computational cost results in little benefit (in terms of 
overall efficiency) over using $N=1$. A special case of aCPMMH (when $\rho=0$ and $N=1$) is the modified 
innovation scheme of \cite{GoliWilk08}, which is typically outperformed (in terms of overall efficiency) with $\rho>0$. 

\subsection{Limitations and future directions}

{There are some limitations of aCPMMH which form the basis of future research. For example, in each of 
our applications the associated latent process exhibits unimodal marginal distributions and linear dynamics 
between observation instants that are well approximated by the modified diffusion bridge construct. Extension to the 
multimodal case would require an importance proposal that captures the multimodality of the marginal distributions 
of the true bridge between observation instants. We refer the reader to the guided proposals of 
\cite{Meulen17} and \cite{schauer17} for possible candidate proposal processes.} 

{Additional directions for future research include the use of methods based on adaptive proposals which may be of 
benefit in both the parameter and latent state update steps. Proposal mechanisms that exploit gradient 
information such as the Hamiltonian Monte Carlo (HMC) method \citep{duane87} may also be of interest. 
We note that in the case of $N=1$, it is possible to directly calculate the required 
log-likelihood gradient. For $N>1$, importance samples generated from $\pi(x^L |x^o, \theta)$ 
could be used to estimate}
\[
\nabla_{\theta}\log \pi(x^o|\theta) = \int \nabla_{\theta} \log \pi(x|\theta) \pi(x^L |x^o, \theta)dx^L .
\]
{For a general discussion on the use of particle filters for estimating log-likelihood gradients we refer 
the reader to \cite{poyiadjis11}; see also \cite{nemeth16}. A comparison of aCPMMH with approaches that 
target the joint posterior of the parameters and latent process also warrants further attention; see e.g. 
\cite{botha20} for an implmentation of the latter.}

Although we have focussed on updating the latent states in separate blocks 
(single site updating), other blocking schemes may offer improved mixing efficiency. Alternatively, 
it might be possible to reduce the number of latent variables, for example, by only explicitly 
including latent states in the joint posterior at every (say) $k$th observation instant. The success 
of such a scheme is likely to depend on the accuracy of an importance sampler that covers $k$ observations, 
and whether or not the resulting likelihood estimates can be made sufficiently correlated. This is the subject 
of ongoing work.        

\appendix

\section{Linear noise approximation (LNA)}\label{lna}

The linear noise approximation (LNA) provides a tractable approximation 
to the SDE in (\ref{eqn:sdmem}). We provide brief, intuitive details of the 
LNA and its implementation, and refer the reader to \cite{fearnhead14} 
(and the references therein) for an in-depth treatment.

\subsection{Derivation and solution}\label{lnaSol}

Partition $X_t$ as 
\begin{equation}\label{eqn:lna_x_part}
X_t=\eta_t+ R_t,
\end{equation} 
where $\{\eta_t, t\geq0\}$ is a deterministic process satisfying 
the ODE
\begin{equation}\label{eqn:lna_eta}
\frac{d\eta_t}{dt} = \alpha(\eta_t,\theta),\qquad \eta_0=x_0
\end{equation} 
and $\{R_t, t\geq0\}$ is a residual stochastic process. The residual 
process $(R_t)$ satisfies
\begin{equation}\label{eqn:resid_sde}
dR_t=\{\alpha(X_t,\theta)-\alpha(\eta_t,\theta)\}\,dt+\sqrt{\beta(X_t,\theta)}\,dW_t
\end{equation}
which will typically be intractable. A tractable 
approximation can be obtained by Taylor expanding $\alpha(X_t,\theta)$ 
and $\beta(X_t,\theta)$ about $\eta_t$. Retaining the first two terms in the 
expansion of $\alpha$ and the first term in the expansion of $\beta$ gives
\begin{equation}\label{eqn:approx_resid_sde}
d\hat{R}_t=H_t\hat{R}_t\,dt+\sqrt{\beta(\eta_t,\theta)}\,dW_t
\end{equation}
where $H_t$ is the Jacobian matrix with ($i$,$j$)th element 
\begin{equation}\label{eqn:jac}
(H_t)_{i,j} = \frac{\partial\alpha_i(\eta_t,\theta)}{\partial\eta_{j,t}}.
\end{equation}
The motivation for the LNA is an underlying assumption that $||X_t-\eta_t||$ 
is ``small'', or in other words, that the drift term 
$\alpha(X_t,\theta)$ dominates the diffusion coefficient 
$\beta(X_t,\theta)$. 

Given an initial condition $\hat{R}_0\sim N(\hat{r}_0,\hat{V}_0)$, we obtain 
$\hat{R}_t$ as a Gaussian random variable. The solution reuires the $d\times d$ fundamental matrix 
$P_t$ that satisfies the ODE 
\begin{equation}\label{eqn:lna_P}
\frac{dP_t}{dt}=H_tP_t,\qquad P_0=I_d, 
\end{equation}
where $I_d$ is the $d\times d$ identity matrix. Now let $U_t=P_t^{-1}\hat{R}_t$ and apply the It\^o formula 
to obtain 
\begin{align*}
dU_t&=P_t^{-1}\sqrt{\beta(\eta_t,\theta)}\,dW_t.
\end{align*}
Hence, we may write
\[
U_t=U_0+\int_{0}^tP_s^{-1}\sqrt{\beta(\eta_s,\theta)}\,dW_s.
\]
Appealing to linearity and It\^{o} isometry we obtain
\begin{equation}\label{eqn:u_norm}
U_t\vert U_0\sim N\left\{U_0,\int_{0}^tP_s^{-1}\beta(\eta_s,\theta)\left(P_s^{-1}\right)^T ds\right\}.
\end{equation}
Therefore, for the initial condition above, we have that 
\[
\hat{R}_t\vert\hat{R}_0=\hat{r}_0\sim N\left(P_t\hat{r}_0,P_t\psi_tP_t^T \right),
\]
where
\[
\psi_t=\hat{V}_0 + \int_{0}^tP_s^{-1}\beta(\eta_s,\theta)\left(P_s^{-1}\right)^T ds.
\]
Setting $m_t=P_t\hat{r}_0$ and $V_t=P_t\psi_tP_t^T $ gives
\[
X_t|X_0 \sim N\left(\eta_t+m_t,V_t\right)
\]
where $\eta_t$, $m_t$ and $V_t$ satisfy the coupled ODE system consisting 
of (\ref{eqn:lna_eta}) and 
\begin{align}
\frac{dm_t}{dt}&=H_tm_t, \qquad m_0=\hat{r}_0,   \label{eqn:lna_m}\\
\frac{dV_t}{dt}&=V_tH_t^T +\beta(\eta_t,\theta)+H_tV_t, \qquad V_0=0.\label{eqn:lna_V}
\end{align}
In the absence of an analytic solution, this system of coupled ODEs
must be solved numerically. Note that if $\eta_0=x_0$ so that $\hat{r}_0=0$, 
$m_t=0$ for all times $t\geq 0$ and (\ref{eqn:lna_m}) need not be solved. 

\subsection{Inference using the LNA}\label{lnaInf}

Consider the LNA as an inferential model. The posterior over parameters 
and the latent process (at the observation times) is denoted by 
${\pi}^{(a)}(\theta,x^o|y)$. We sample this posterior in two steps. Firstly, 
a Metropolis-Hastings scheme is used to target the marginal parameter posterior
\begin{align}
{\pi}^{(a)}(\theta|y)&\propto \pi(\theta){p}^{(a)}(y|\theta) \label{postLNA}
\end{align}
where ${p}^{(a)}(y|\theta)$ is the marginal likelihood under the LNA. Then, 
a sample $x^o$ is drawn from $p^{(a)}(x^o|\theta,y)$ for each $\theta$ sample 
from step one. Note that $p^{(a)}(y|\theta)$ and $p^{(a)}(x^o|\theta,y)$ are tractable under the LNA. 

A forward filter is used to evaluate $p^{(a)}(y|\theta)$. Since the parameters $\theta$ remain fixed throughout 
the calculation of ${p}^{(a)}(y|\theta)$, we drop them from
the notation where possible. 

Define $y_{1:t}=(y_{1},\ldots,y_{t})^T$. It will additionally be helpful 
to adopt the notational convention that ${p}^{(a)}(y_{1}|y_{1:0})={p}^{(a)}(y_{1})$ 
and set ${p}^{(a)}(y_{1:0})=1$. Now suppose that $X_0\sim N(a_0,C_0)$ \emph{a priori}.
The marginal likelihood ${p}^{(a)}(y|\theta)$ under the LNA can be obtained 
using Algorithm~\ref{alg:lna_ffbs}. 

\begin{algorithm}[t!]
\caption{LNA forward filter} \label{alg:lna_ffbs}
\begin{enumerate}
\item For $t=0,\ldots,n-1$,
\begin{itemize}
\item[(a)] Prior at $t+1$. Initialise the LNA with $\eta_{t}=a_{t}$, 
$V_{t}=C_{t}$ and $P_{t}=I_d$. Integrate the ODEs 
\eqref{eqn:lna_eta}, \eqref{eqn:lna_V} and \eqref{eqn:lna_P} forward to $t+1$ to obtain 
$\eta_{t+1}$, $V_{t+1}$ and $P_{t+1}$. 
\item[(b)] One step forecast. Using the observation equation \eqref{eqn:obs}, 
we have that 
\[
Y_{t+1}|y_{1:t}\sim N\left(F^T \eta_{t+1},F^T V_{t+1}F+\Sigma\right).
\]
Compute the updated marginal likelihood
\begin{align*}
{p}^{(a)}(y_{1:t+1})&={p}^{(a)}(y_{1:t}){p}^{(a)}(y_{t+1}|y_{1:t}) \\
&={p}^{(a)}(y_{1:t}) \\
& \, \times N\left(y_{t+1}\,;\, F^T \eta_{t+1}\,,\,F^T V_{t+1}F+\Sigma\right).
\end{align*}
\item[(c)] Posterior at $t+1$. Combining the distributions in (a) and (b) gives the joint 
distribution of $X_{t+1}$ and $Y_{t+1}$ (conditional on $y_{1:t}$) as
\[
\begin{pmatrix}
	X_{t+1} \\	
	Y_{t+1}
	\end{pmatrix}\sim N\left\{\begin{pmatrix}
	\eta_{t+1} \\[0.2em]	
	F^T \eta_{t+1} 	
	\end{pmatrix}\,,\, \begin{pmatrix}
	V_{t+1} & V_{t+1}F  \\[0.2em]	
	F^T V_{t+1} & F^T V_{t+1}F+\Sigma  	 
	\end{pmatrix} \right \} 
\]
and therefore $X_{t+1}|y_{1:t+1}\sim N(a_{t+1},C_{t+1})$, where
\begin{align*}
a_{t+1} &= \eta_{t+1}+V_{t+1}F\left(F^T V_{t+1}F+\Sigma\right)^{-1}\\
& \, \times \left(y_{t+1}-F^T \eta_{t+1}\right) \\
C_{t+1} &= V_{t+1}-V_{t+1}F\left(F^T V_{t+1}F+\Sigma\right)^{-1}F^T V_{t+1}\,.
\end{align*}
Store the values of $a_{t+1}$, $C_{t+1}$, $\eta_{t+1}$, $V_{t+1}$ and $P_{t+1}$.
\end{itemize}
\end{enumerate}
\end{algorithm}

Hence, samples of $\theta$ can be obtained from ${\pi}^{(a)}(\theta|y)$ by running a Metropolis-Hastings scheme 
with target (\ref{postLNA}). Then, for each (thinned) $\theta$ draw, $x^o$ can be sampled from 
$p^{(a)}(x^o|\theta,y)$ using a backward sampler (see Algorithm~\ref{alg:lna_bs}). 
 
\begin{algorithm}[t!]
\caption{LNA backward sampler} \label{alg:lna_bs}
\begin{enumerate}
\item First draw $x_{n}$ from $X_{n}|y\sim
  N(a_{n},C_{n})$.
\item For $t=n-1,n-2,\ldots,1$,
\begin{itemize}
\item[(a)] Joint distribution of $X_{t}$ and $X_{t+1}$. Note
  that $X_{t}|y_{1:t}\sim N(a_{t},C_{t})$. The joint
  distribution of $X_{t}$ and $X_{t+1}$ (conditional on
  $y_{1:t}$) is 
\[
\begin{pmatrix}
	X_{t} \\	
	X_{t+1}
	\end{pmatrix}\sim N\left\{\begin{pmatrix}
	a_{t} \\[0.2em]	
	\eta_{t+1} 	
	\end{pmatrix}\,,\, \begin{pmatrix}
	C_{t} & C_{t}P_{t+1}^T   \\[0.2em]	
	P_{t+1}C_{t} & V_{t+1}  	 
	\end{pmatrix} \right \}. 
\]
\item[(b)] Backward distribution. The distribution of
  $X_{t}|X_{t+1},y_{1:t}$ is
  $N(\hat{a}_{t},\hat{C}_{t})$, where
\begin{align*}
\hat{a}_{t} &= a_{t}+C_{t}P_{t+1}^T V_{t+1}^{-1}\left(x_{t+1}-\eta_{t+1}\right), \\
\hat{C}_{t} &= C_{t}-C_{t}P_{t+1}^T V_{t+1}^{-1}P_{t+1}C_{t}.
\end{align*}
Draw $x_{t}$ from $X_{t}|X_{t+1},y_{1:t}\sim
N(\hat{a}_{t},\hat{C}_{t})$.
\end{itemize} 
\end{enumerate}
\end{algorithm}

\bibliographystyle{apalike}
\bibliography{bridgebib}

\end{document}